\documentclass[11pt]{article}
\pdfoutput=1
\usepackage{graphicx}
\usepackage{hyperref}
\usepackage{epsfig}
\usepackage{amsmath}
\usepackage{amsthm}
\usepackage{amssymb}
\usepackage{multirow}
\usepackage{color}
\usepackage[nosort]{cite}
\usepackage{hyperref}
\usepackage[fontsize=11.45pt]{scrextend}
\usepackage{braket,mathrsfs,slashed}
\usepackage{float}
\usepackage{setspace}
\usepackage[caption=false]{subfig}
\newlength{\abstractwidth}
\newcommand{\be}{\begin{equation}}
\newcommand{\ee}{\end{equation}}

\renewcommand{\title}[1]{\vbox{\center\bf{\Large{#1}}}\vspace{5mm}}
\renewcommand{\author}[1]{\vbox{\center#1}\vspace{5mm}}
\newcommand{\address}[1]{\vbox{\center\em#1}}

\usepackage{dsfont}

\usepackage[utf8]{inputenc}
\usepackage{rotating}
\usepackage{tikz-feynman}

\usepackage{dsfont}
\usepackage[paper=a4paper,margin=1.5cm]{geometry}
\usepackage[utf8]{inputenc}
\usepackage{rotating}
\usepackage{soul}

\usepackage{mathrsfs} 
\usepackage{bbm}
\usepackage{etoolbox} 
\usepackage{multirow}

\allowdisplaybreaks

\renewcommand\[{\begin{equation}}
\renewcommand\]{\end{equation}}

\newcommand{\ba}{\begin{eqnarray}}
\newcommand{\ea}{\end{eqnarray}}

\hypersetup{pdfstartview=FitV,colorlinks=true,linkcolor=midblue,citecolor=midblue,filecolor=midblue,urlcolor=midblue}

\definecolor{midblue}{rgb}{0,0,0.5}

\begin{document}
	
		\newgeometry{top=3.1cm,bottom=3.1cm,right=2.4cm,left=2.4cm}
		
	\begin{titlepage}
	\begin{center}
		\hfill \\
		\vskip 0.5cm

		\title{Dust collapse and horizon formation\\[2mm] in Quadratic Gravity}

			\author{\large Luca Buoninfante$^{a,\,\star}$, Francesco Di Filippo$^{b,\,\dagger}$,\\[1.5mm] Ivan Kolář$^{b\,\ddagger}$, Frank~Saueressig$^{a\,,\diamond}$
			 }
			
			\address{$^a$High Energy Physics Department, Institute for Mathematics, Astrophysics,\\
			and Particle Physics, Radboud University, Nijmegen, The Netherlands\\[1.5mm]
				$^b$Institute of Theoretical Physics, Faculty of Mathematics and Physics, Charles\\ University, V.~Holešovičkách 2, 180 00 Prague 8, Czech Republic}
				\vspace{.3cm}

		\end{center}

\vspace{0.15cm}

\begin{abstract}
Quadratic Gravity supplements the Einstein-Hilbert action by terms quadratic in the spacetime curvature. This leads to a rich phase space of static, compact gravitating objects including the Schwarzschild black hole, wormholes, and naked singularities. For the first time, we study the collapse of a spherically symmetric star with uniform dust density in this setting. We assume that the interior geometry respects the symmetries of the matter configuration, i.e., homogeneity and isotropy, thus it is insensitive to the Weyl-squared term and the interior dynamics is fully determined by $R$ and $R^2$. 
As our main result, we find that the collapse leads to the formation of a horizon, implying that the endpoint of a uniform dust collapse with a homogeneous and isotropic interior is not a horizonless spacetime. We also show that the curvature-squared contribution is responsible for making the collapse into a singularity faster than the standard Oppenheimer-Snyder scenario. Furthermore, the junction conditions connecting spacetime inside and outside the matter distribution are found to be significantly more constraining than their counterparts in General Relativity and we discuss key properties of any exterior solution matching to the spacetime inside the collapsing star. Finally, we comment on the potentially non-generic behavior entailed by our assumptions. 
\end{abstract}
\vspace{5cm}
\noindent\rule{6.5cm}{0.4pt}\\
$\,^\star$
\href{mailto:luca.buoninfante@ru.nl}{luca.buoninfante@ru.nl}\\	
$\,^\dagger$ \href{mailto:francesco.difilippo@matfyz.cuni.cz}{francesco.difilippo@matfyz.cuni.cz}\\
$\,^\ddagger$ \href{mailto:ivan.kolar@matfyz.cuni.cz}{ivan.kolar@matfyz.cuni.cz}
\\
$\,^\diamond$ \href{mailto:f.saueressig@science.ru.nl}{f.saueressig@science.ru.nl}

\end{titlepage}

{
	\hypersetup{linkcolor=black}
	\tableofcontents
}

\baselineskip=17.63pt



\newpage

\section{Introduction}

The addition of quadratic-curvature terms to the Einstein-Hilbert action gives rise to a perturbatively strictly renormalizable local quantum field theory of gravity in four spacetime dimensions~\cite{Stelle:1976gc} (see Refs.~\cite{Salvio:2018crh,Donoghue:2021cza,Piva:2023bcf,Buoninfante:2022ykf} for reviews). This theory is known as Quadratic Gravity and its action reads 
\begin{equation}
    S[g]=\frac{1}{2}\int {\rm d}^4x \sqrt{-g}\left[\frac{1}{8\pi G_{\rm N}}R+\frac{\alpha}{6} R^2 -\frac{\beta}{2}C_{\mu \nu \rho \sigma} C^{\mu \nu \rho \sigma}\right]\,,
    \label{action}
\end{equation}
where $G_{\rm N}$ is Newton's constant, and $\alpha$ and $\beta$ are dimensionless parameters. We have neglected the contribution of the cosmological constant and the boundary terms including the Gauss-Bonnet which in four dimensions is topological. 

The quadratic-curvature terms introduce extra degrees of freedom in addition to the massless graviton: a massive spin-$0$ of mass squared $m_0^2=\frac{1}{8\pi G_{\rm N}}\frac{1}{\alpha}$ coming from $R^2$ and a massive spin-$2$ of mass squared\footnote{It is worthwhile mentioning that the mass of the spin-$2$ depends on the cosmological constant $\Lambda$. In fact, when $\Lambda\neq 0$ in four spacetime dimensions we have $m^2_2=\frac{1}{8\pi G_{\rm N}}\frac{1}{\beta}+\frac{2}{3}\Lambda(2\frac{\alpha}{\beta}+1)$~\cite{Anselmi:2018tmf,Buoninfante:2023ryt}.} $m_2^2=\frac{1}{8\pi G_{\rm N}}\frac{1}{\beta}$ coming from $C_{\mu\nu\rho\sigma}C^{\mu\nu\rho\sigma}.$ Note that $\alpha$ and $\beta$ must be positive in order to avoid tachyons.\footnote{Notably, Lagrangians of the form $R+R^2$ are very successful when explaining the power spectrum of the cosmic microwave background (CMB) via Starobinsky inflation~\cite{Starobinsky:1980te,Starobinsky:1983zz,Planck:2018jri}. For this mechanism to work, the spin-$0$ mass is required to be of order $m_0\sim 10^{13}$GeV which also means $1/m_0\sim 10^{-30}$m. The scale $1/m_2$ is expected to be of the same order or smaller.}
The massive spin-$2$ field is ghost-like due to unconstrained higher-order time derivatives, meaning that its individual kinetic term (or, in other words, its propagator) is multiplied by an overall minus sign as compared to standard two-derivative fields. On the other hand, the spin-$0$ component is subject to the Hamiltonian constraint and the kinetic term of the additional scalar degree of freedom has the standard positive sign.

The presence of the spin-$2$ ghost in the particle spectrum could raise concerns about the feasibility of the theory due to potential classical Hamiltonian instabilities and violation of unitarity at the quantum level~\cite{Stelle:1976gc,Woodard:2015zca}. However, especially in the recent years, new promising proposals have emerged on how to quantize ghost fields consistently with the principles of locality, unitarity and renormalizability~\cite{Salvio:2014soa,Anselmi:2017ygm,Anselmi:2018ibi,Anselmi:2018bra,Donoghue:2019fcb,Donoghue:2019ecz,Holdom:2021hlo}, and also new ideas suggesting that ghosts may not always lead to classical instabilities~\cite{Deffayet:2021nnt,Deffayet:2023wdg,ErrastiDiez:2024hfq}. Here we do not discuss the ghost puzzle and its possible solution(s) as this topic would go beyond the scope of the paper. Moreover, our analysis will be confined to the purely classical level without taking quantum corrections into account.

\subsubsection*{Aim of this work} 

We are interested in studying the properties of asymptotically flat and spherically symmetric metrics that can result as physical endpoints of a gravitational collapse in Quadratic Gravity. Most of the previous works only focused on finding \textit{static} spherically symmetric solutions which were classified in terms of the asymptotic behaviors of the two metric components $g_{tt}(r)$ and $g_{rr}(r)$ (in Schwarzschild coordinates). Different types of static solutions have been  found~\cite{Stelle:1977ry,Holdom:2002xy,Lu:2015psa,Lu:2015cqa,Perkins:2016imn,Lu:2017kzi,Holdom:2016nek,Podolsky:2018pfe,Podolsky:2019gro,Bonanno:2019rsq,Bonanno:2021zoy,Daas:2022iid,Bonanno:2022ibv,Silveravalle:2022wij,Huang:2022urr}: singular black hole solutions (with horizon radius $r_H>0$ such that $g_{tt}(r_H)=0$) of Schwarzschild and non-Schwarzschild type, horizonless solutions with naked singularities, and wormholes. In particular, all the static, asymptotically flat, spherically symmetric black-hole solutions can be shown to have zero Ricci scalar, $R=0$~\cite{Nelson:2010ig,Lu:2015cqa}.

It is also worth to mention that stability of time-dependent perturbations of Ricci flat solutions in Quadratic Gravity have been performed in~\cite{Held:2022abx,Hell:2023mph} and numerical-relativity methods have begun to be successfully implemented for quadratic-curvature Lagrangians~\cite{Held:2023aap,East:2023nsk,Cayuso:2023dei}. In particular, the initial value problem in Quadratic Gravity is well-posed~\cite{Noakes:1983xd,Figueras:2024bba}.

While observations based on shadow imaging~\cite{EventHorizonTelescope:2019dse,EventHorizonTelescope:2022wkp} allow to rule out parts of the static-solutions space~\cite{Daas:2022iid}, it is still unclear whether the more exotic naked singularities and wormhole solutions are physical or not. In other words, it has not yet been proven which of these solutions can form as a result of a dynamical process, that is, whether they can be the physical endpoint of a gravitational collapse. For example, it could be that some or even all of them cannot be matched to a dynamical process.

In this paper we aim at addressing this physical question for the first time\footnote{Let us mention that Ref.~\cite{Cayuso:2023dei} studied the gravitational collapse of a massless scalar field for an action containing quadratic curvatures but in the framework of effective field theory where no additional gravitational degree of freedom is present besides the massless graviton and the higher-curvature terms can be partly removed and partly moved into the matter sector through a field redefinition. In our case, instead, we work in the renormalizable theory of Quadratic Gravity in which $R^2$ and $C_{\mu\nu\rho\sigma}C^{\mu\nu\rho\sigma}$ cannot be removed as they are not corrections but non-perturbative modifications of Einstein's theory.}. The main goal is to describe a dynamical scenario and understand the properties of the resulting spacetime. We consider the simplest possible situation in which the matter configuration undergoing gravitational collapse is a uniform ball of dust\footnote{Uniform dust collapse has also been recently studied in an asymptotic-safety inspired scenario where it was found that the endpoint of the collapse is a regular black hole~\cite{Bonanno:2023rzk}.} whose interior spacetime is homogeneous and isotropic as originally done by Oppenheimer and Snyder in their seminal work on gravitational singularities~\cite{Oppenheimer:1939ue}. In particular, we make the reasonable physical assumption according to which in the initial phase of the collapse General Relativity (GR) and known standard physics work well, and analyse the effect of the higher curvature terms on the collapse. We divide the analysis into interior and exterior solutions and organize the work as follows.
\begin{description}
   
    \item[\textbf{Sec.~\ref{sec:quad-grav}:}] We recall the Quadratic Gravity field equations and briefly review the families of static, asymptotically flat, spherically symmetric solutions that were found in the past.
    
    \item[\textbf{Sec.~\ref{sec:unif-collap}:}] We assume that the interior geometry respects the same symmetries of the matter configuration, i.e., homogeneity and isotropy. This implies that the interior solution is insensitive to the Weyl term, therefore the interior dynamics is solely determined by $R$ and $R^2$ in the action. We show that the collapse into a singularity is faster than in the case of GR.
    
    \item[\textbf{Sec.~\ref{sec:horiz-form}:}] We show that an apparent horizon forms by analysing the evolution of the expansion parameters of the two vectors of a null geodesics congruence, that become both negative at some finite time during the collapse. 

    \item[\textbf{Sec.~\ref{sec:exter}:}] Subsequently, we consider the exterior solution by studying the junction conditions and comment on how to constrain the two metric functions $g_{tt}(t,r)$ and $g_{rr}(t,r).$ We show that no stationary solution  that smoothly matches a collapsing uniform-density dust star can be found. This means that at a generic time the exterior metric components must be time-dependent and that stationarity (i.e. staticity in the spherically symmetric case) can only be reached asymptotically. In this paper we do not attempt to find an explicit exterior solution but for the time being we only constrain some of its properties. In particular, our results imply that horizonless spherically symmetric solutions in Quadratic Gravity cannot be matched to a dynamical uniform-density dust configuration. This suggests that some of the known static solutions, such as the 2-2 hole~\cite{Holdom:2016nek}, cannot be the endpoint of a uniform-density dust collapse.

    \item[\textbf{Sec.~\ref{sec:conclus}}] We summarize the main results, make some final remarks on the singularity problem and draw the conclusions. In particular, we note that the Weyl squared term could become important in the interior in less symmetric scenarios (e.g. in the presence of inhomogeneities, anisotropies and/or rotation) and contribute with a repulsive contribution that may affect the collapse in non-trivial ways.
    
\end{description}

\paragraph{Conventions and notations.} We adopt the mostly plus convention for the metric signature ($-+++$) and work in Natural units $c=1=\hbar$. We choose the following convention for the Riemann tensor: $[\nabla_\nu,\nabla_\rho]V^\sigma=V^\mu {R^\sigma}_{\mu\nu\rho},$ ${R^{\sigma}}_{\mu\nu\rho}=\partial_\nu{\Gamma^\sigma}_{\mu\rho}-\partial_\rho{\Gamma^\sigma}_{\mu\nu}+{\Gamma^\sigma}_{\alpha\nu}{\Gamma^\alpha}_{\mu\rho}-{\Gamma^\sigma}_{\alpha\rho}{\Gamma^\alpha}_{\mu\nu},$  and $R_{\mu\rho}={R^\nu}_{\mu\nu\rho}=\delta^{\nu}_\sigma {R^{\sigma}}_{\mu\nu\rho}=g^{\sigma\nu}R_{\sigma\mu\nu\rho}.$ Since we will consider the interior and exterior spacetime regions of a collapsing ball of dust, we will distinguish these regions with the labels “$+$” and “$-$”. For instance, the metric components in the interior and exterior regions will be called $g^-_{\mu\nu}(x_-)$ and $g^+_{\mu\nu}(x_+),$ where $x_-^\mu=(\tau,\chi,\theta,\varphi)$ and $x_+^\mu=(t,r,\theta,\varphi)$ are the interior and exterior coordinates, respectively.

\section{Quadratic Gravity}\label{sec:quad-grav}

\subsection{Field equations}

In four spacetime dimensions we can use the relation $C_{\mu\nu\rho\sigma}C^{\mu\nu\rho\sigma}=R_{\mu\nu\rho\sigma}R^{\mu\nu\rho\sigma}-2R_{\mu\nu}R^{\mu\nu}+\frac{2}{3}R^2$ and the fact that the Gauss-Bonnet density $\sqrt{-g}(R_{\mu\nu\rho\sigma}R^{\mu\nu\rho\sigma}-4R_{\mu\nu}R^{\mu\nu}+R^2)$ is locally a total derivative to rewrite the Weyl-squared invariant in the action in terms of the squares of the Ricci scalar and Ricci tensor, thus up to total derivatives eq.~\eqref{action} can be recast as 
\begin{equation}
	S[g]=\frac{1}{2}\int {\rm d}^4x\sqrt{-g}\left[\frac{1}{8\pi G_{\rm N}}R+\frac{\alpha}{6}R^2-\beta\left(R_{\mu\nu}R^{\mu\nu}-\frac{1}{3}R^2\right)\right]\,.\label{action-weyl->Ricci}
\end{equation}
We can derive the field equations by varying the action in~\eqref{action-weyl->Ricci} and obtain~\cite{Stelle:1977ry}
\begin{equation}
E_{\mu\nu}=T_{\mu\nu}\,,
\label{eom}
\end{equation}
where\footnote{Note that starting from the form of the action written in terms of the Weyl square, i.e. eq.~\eqref{action}, the contribution proportional to $\beta$ in the field equations~\eqref{lhs-eom} would read $-2\beta B_{\mu\nu}\equiv -2\beta(\nabla^\rho\nabla^\sigma+\frac{1}{2}R^{\rho\sigma})C_{\mu\rho\nu\sigma},$ where $B_{\mu\nu}$ is known as Bach tensor and is trace-free. In four spacetime dimensions the Bach tensor can be expressed in terms of Ricci scalar and Ricci tensor (and no Riemann tensor) as $B_{\mu\nu}=\frac{1}{2}\Box R_{\mu\nu}+\frac{1}{6}(2\nabla_\mu\nabla_\nu-\frac{1}{2}g_{\mu\nu}\Box)R-\frac{1}{2}\nabla_{\rho}\nabla_\mu R_\nu^\rho-\frac{1}{2}\nabla_{\rho}\nabla_\nu R_\mu^\rho-\frac{1}{3}RR_{\mu\nu}+R_{\mu\rho}R^\rho_\nu-\frac{1}{4}(R^{\rho\sigma}R_{\rho\sigma}-\frac{1}{3}R^2)g_{\mu\nu},$ consistently with eq.~\eqref{lhs-eom}.}
\begin{eqnarray}
E_{\mu\nu}&\equiv&\frac{1}{8\pi G_{\rm N}}\left(R_{\mu\nu}-\frac{1}{2}g_{\mu\nu}R\right)+\frac{2}{3}\left(\frac{\alpha}{2}+\beta\right)\left(g_{\mu\nu}\Box R-\nabla_\mu\nabla_\nu R+R R_{\mu\nu}-\frac{1}{4}g_{\mu\nu} R^2\right) \nonumber\\[2mm]
&&-\beta\left(\Box R_{\mu\nu} +\frac{1}{2}g_{\mu\nu}\Box R-\nabla_\rho \nabla_{\mu}R^\rho_{\nu}-\nabla_\rho \nabla_{\nu}R^\rho_{\mu} +2R_{\mu}^\rho R_{\rho\nu}-\frac{1}{2}g_{\mu\nu}R_{\rho\sigma}R^{\rho\sigma} \right) \,,
\label{lhs-eom}
\end{eqnarray}
and $T_{\mu\nu}$ is the stress-energy tensor of the matter sector, which is covariantly conserved. The trace of the field equations reads
\begin{eqnarray}
E=-\frac{1}{8\pi G_{\rm N}}R+\alpha \Box R = T\,,
\label{trace-eom}
\end{eqnarray}
where we have defined $E\equiv g^{\mu\nu}E_{\mu\nu}$ and $T\equiv g^{\mu\nu}T_{\mu\nu}.$
As expected, the terms proportional to $\beta$ do not contribute to the trace equation because the Weyl tensor is trace-free.

\subsection{Brief summary of static solutions}\label{sec:summ-static-sol}

The field equations~\eqref{eom},~\eqref{lhs-eom} have been solved in the case of a static and spherically symmetric metric ansatz of the type~\cite{Stelle:1977ry,Holdom:2002xy,Lu:2015psa,Lu:2015cqa,Lu:2017kzi,Holdom:2016nek,Podolsky:2019gro}
\begin{eqnarray}
ds^2=-A(r){\rm d}t^2+B(r){\rm d}r^2+r^2({\rm d}\theta^2+\sin^2\theta {\rm d}\varphi^2)\,,
\label{ext-ansatz}
\end{eqnarray}
where $A(r)$ and $B(r)$ are two (in general, independent) unknown functions.
Due to the more complicated structure of the field equations as compared to Einstein's theory it is not possible to find non-linear solutions in closed form. However, approximate analytic solutions can be found by working in the linear regime when the gravitational potentials are weak, e.g. at large distances, and also by expanding around some spacetime point that can be either the origin or some finite $r_0>0.$
We now briefly review the main properties of these solutions.

\paragraph{Linearized solutions.} In the large-distance regime we expect the gravitational field to be weak and an approximate analytic solution can be obtained by solving the linearized equations. The general solution contains six free parameters and is given by~\cite{Stelle:1977ry,Bonanno:2021zoy}
\begin{eqnarray}
A(r)&\simeq & 1+C_t-\frac{2G_{\rm N}M}{r}+2S_2^+ \frac{e^{m_2r}}{r}+2S_2^- \frac{e^{-m_2r}}{r}+S_0^+\frac{e^{m_0r}}{r}+S_0^-\frac{e^{-m_0r}}{r}\,, \nonumber \\[1.5mm]
B^{-1}(r)&\simeq& 1-\frac{2G_{\rm N}M}{r}+S_2^+(1-m_2r)\frac{e^{m_2r}}{r}+S_2^- (1+m_2r)\frac{e^{-m_2r}}{r}\nonumber\\
&&-S_0^+(1-m_0r)\frac{e^{m_0r}}{r}-S_0^-(1+m_0r)\frac{e^{-m_0r}}{r}\,,
\label{lin-sol}
\end{eqnarray}
where $C_t,$ $M,$ $S_2^+$, $S_2^-$, $S_0^+$ and $S_0^-$ are six integration constants that form a six-dimensional space of solutions. The number of free parameters can be reduced by requiring asymptotic flatness, i.e., $S_2^+=0=S_0^+$, and choosing a canonical parametrization for the time coordinate, i.e., $C_t=0.$

\paragraph{Expansion around the origin.} The idea is to assume a Laurent series expansion for the metric potentials, 
\begin{eqnarray}
A(r)=r^t\sum\limits_{n=0}^\infty a_n r^n\,,\qquad B(r)=r^s\sum\limits_{n=0}^\infty b_n r^n\,,
\label{metric-frobenius}
\end{eqnarray}
where $a_0\neq 0,$ $b_0\neq 0,$ and determine the leading order behavior, i.e. the values of $t$ and $s;$ this procedure is also known as Frobenius method.

In Ref.~\cite{Stelle:1977ry} three families of solutions were found:
\begin{itemize}

\item $(s,t)=(0,0)$: it is the natural vacuum solution that is regular everywhere and is characterized by three independent parameters; 

\item $(s,t)=(1,-1)$: it contains spacetime metrics with Schwarzschild-like singularities for which the Kretschmann invariant diverges like $R^{\mu\nu\rho\sigma}R_{\mu\nu\rho\sigma}\sim 1/r^6,$ and is characterized by four independent parameters;

\item $(s,t)=(2,2)$: it contains other type of singular spacetime metrics whose Kretschmann invariant diverges like $R^{\mu\nu\rho\sigma}R_{\mu\nu\rho\sigma}\sim 1/r^8,$ and is characterized by six independent parameters.

\end{itemize}
Note that one parameter in $A(r)$ always corresponds to the rescaling of the time coordinate. Further parameters can be fixed by extra conditions, e.g., $R=0$ fixes one more and, moreover, the asymptotic flatness can be achieved for special values of the parameters.

\paragraph{Exansion around $r_0>0.$} The above classification was extended in Refs.~\cite{Lu:2015cqa,Lu:2015psa} by applying the Frobenius method to expansions around a generic finite radius $r_0>0:$
\begin{eqnarray}
A(r)=(r-r_0)^t\sum\limits_{n=0}^\infty a_n (r-r_0)^n\,,\qquad B(r)=(r-r_0)^s\sum\limits_{n=0}^\infty b_n (r-r_0)^n\,.
\label{metric-frobenius}
\end{eqnarray}
This more general analysis allows to clarify whether the singular solutions above have a horizon or a naked singularity. The solution families that were found are~\cite{Lu:2015psa} 
\begin{itemize}

\item $(s,t)_{r_0}=(0,0)_{r_0}:$ this is the regular branch, this time with a generic  expansion point $r_0;$ 

\item $(s,t)_{r_0}=(-1,1)_{r_0}:$ these are black hole solutions where $r_0$ is the location of the horizon;

\item $(s,t)_{r_0}=(-1,0)_{r_0}:$ these solutions describe wormholes whose throat is given by $r_0$.

\end{itemize}
Furthermore, it is also worth to mention that there exist other non-Frobenius solutions in these coordinates~\cite{Lu:2015psa} that are Frobenius in conformal-to-Kundt coordinates (for $R=0$)~\cite{Podolsky:2019gro}.

The class $(-1,1)_{r_0}$ contains the Schwarzschild black hole with $b_0=r_0$ as an isolated solution, i.e. small deviations from it do not give rise to new solutions, and an additional non-Schwarzschild black hole solution also known as Schwarzschild-Bach black hole.  The latter is characterized by an extra (in general not small) parameter $\delta,$ defined through the relation $b_0=r_0/(1+\delta),$ whose non-zero value implies $A(r)\neq 1/B(r).$ The mass parameter of the Schwarzschild-Bach black hole is not arbitrary but bounded from above by $\mathcal{O}\left(M^2_{\rm p}/ m_2\right)$~\cite{Lu:2015cqa}, where $M_{\rm p}=1/\sqrt{8\pi G_{\rm N}}\sim 10^{18}$GeV is the reduced Planck mass.

It is also important to mention that using the trace equation~\eqref{trace-eom} one can prove a no-hair-type theorem stating that all the static, asymptotically flat and spherically symmetric black holes in Quadratic Gravity have Ricci scalar equal to zero~\cite{Nelson:2010ig,Lu:2015cqa}. This implies, for instance, that in $R+R^2$ gravity (i.e., the theory~\eqref{action} with $\beta=0$) the only asymptotically-flat black hole solution is the Schwarzschild metric.

Finally, let us mention that the $(2,2)$ family found expanding around $r=0$ has been shown to be horizonless and, therefore, to have a naked singularity~\cite{Holdom:2002xy,Lu:2015psa}.
This type of solutions are also known as 2-2 holes~\cite{Holdom:2016nek,Holdom:2022zzo}.

\section{Uniform dust collapse}\label{sec:unif-collap}

All the solutions reviewed in the previous section  are static. This means that it is not yet clear whether they can form through a dynamical process, that is, whether they can arise as the physical endpoint of a gravitational collapse. We address this question for the first time by considering the simplest scenario of a collapsing spherically symmetric star of uniform-density dust. In other words, we aim at generalizing the Oppenheimer-Snyder model of collapse~\cite{Oppenheimer:1939ue} to the case of Quadratic Gravity.

We assume that the interior spacetime respects the symmetries of the matter configuration, i.e., homogeneity and isotropy. This implies that the geometry is given by a Friedmann–Lemaitre– Robertson–Walker (FLRW)-type metric which in comoving coordinates $x^\mu_-\equiv (\tau,\chi,\theta,\varphi)$ reads 
\begin{equation}
    ({\rm d}s^2)_{-} =g_{\mu\nu}^-(x_-){\rm d}x_-^\mu{\rm d}x_-^\nu=-{\rm d}\tau^2+a^2(\tau)\left[\frac{{\rm d}\chi^2}{1-k \chi^2}+\chi^2{\rm d}\Omega^2\right]\,,
    \label{interior-ansatz}
\end{equation}
where $a(\tau)$ is the scale factor, $k$  the spatial curvature and we recall that the notation $(\cdot)_-$ refers to quantities defined in the interior region of the collapsing star. 

The matter configuration we consider is a spherically symmetric and pressureless ball of dust with uniform density whose stress-energy tensor reads
\begin{equation}
T_{\mu\nu} = \rho (\tau) U_\mu U_\nu \,,
    \label{stress-energy-dust}
\end{equation}
where $U^\mu=(1,0,0,0)$ is the fluid velocity in  comoving coordinates. 

On the inner side the surface of the collapsing star can be described in terms of the coordinates $x_{-*}^\mu=(\tau,\chi_*,\theta,\varphi)$, where $\chi_*$ is the radius of the star which is a constant in comoving coordinates. Note that the metric~\eqref{interior-ansatz} is valid for $\chi\leq \chi_*.$

The only unknown function that needs to be determined by solving the field equations in the interior is $a(\tau)$, while the constant parameter $k$ will be related to the initial conditions of both matter and geometry configurations, and to the parameters of the exterior geometry through the junction conditions.

\subsection{Interior field equations}

We can insert the ansatz~\eqref{interior-ansatz} and~\eqref{stress-energy-dust}  into the field equations and obtain
\begin{eqnarray}
&&E_{\tau\tau}= \frac{3}{8\pi G_{\rm N}}\frac{1}{a^2}\left(k+\dot{a}^2\right) +3\alpha\left[ \frac{1}{a^4}\left(k^2-2 k \dot{a}^2-3\dot{a}^4 \right)-\frac{\ddot{a}^2}{a^2} +2\frac{\dot{a}\dddot{a}}{a^2}+2\frac{\dot{a}^2\ddot{a}}{a^3}\right] =\rho\,, \label{eomin-00}\\[2mm]
&&E_{\chi\chi}= \frac{-1}{1-k \chi^2}\left\{\frac{1}{8\pi G_{\rm N}}\left(k +\dot{a}^2+2a \ddot{a} \right)\right. \nonumber\\[2mm]
&&\left.\qquad \quad +\alpha \left[3\ddot{a}^2+4\dot{a}\dddot{a}+2a\ddddot{a}- \frac{1}{a^2}\left( k^2-2k\dot{a}^2-3\dot{a}^4 \right)-\frac{4}{a}\left(k+3\dot{a}^2\right)\ddot{a}\right]\right\}= 0\,, \label{eomin-11} \\[2mm]
&&E_{\theta\theta}=\chi^2(1-k\chi^2)E_{\chi\chi}= 0\,, \label{eomin-22}\\[2mm]
&&E_{\varphi\varphi}=\chi^2(1-k\chi^2)\sin^2\theta\,E_{\chi\chi}= 0 \label{eomin-33}\,,
\end{eqnarray}
while the trace reads
\begin{eqnarray}
\!E=-\frac{3}{4\pi G_{\rm N}}\left[ \frac{1}{a^2}\left(k+\dot{a}^2\right)+\frac{\ddot{a}}{a} \right]+6\alpha\left[ \frac{\ddot{a}}{a^3}\left(2k+5\dot{a}^2\right)-\frac{\ddot{a}^2}{a^2}-\frac{1}{a^2}\left( 3\dot{a}\dddot{a}+a\ddddot{a} \right) \right]=-\rho\,,\,
\label{trace-int}
\end{eqnarray}
where the dot notation stands for the derivative with respect to the comoving time $\tau,$ i.e. $\dot{a}(\tau)\equiv \frac{{\rm d}a(\tau)}{{\rm d}\tau}.$
Note that the interior field equations do not depend on $\beta$ because the Weyl tensor vanishes for the FLRW metric ansatz~\eqref{interior-ansatz}, being conformally flat. 

The gravitational field equations have to be compatibly solved with the continuity equation coming from the conservation of the stress-energy tensor, i.e. $\nabla_\mu T^{\mu\nu} = 0,$ from which we obtain
\begin{equation}
3\rho (\tau)\dot{a}(\tau)+\dot{\rho}(\tau) a(\tau)=0 \quad \Leftrightarrow \quad \frac{{\rm d}}{{\rm d}\tau}\left[\rho(\tau) a^3(\tau)\right]=0\,.
    \label{conserv-stress}
\end{equation}
Choosing $a(0)=1$ (always possible by time translation), eq.~\eqref{conserv-stress} gives
\begin{equation}
\rho(\tau)=\rho_0\frac{1}{a^3(\tau)}\,,
    \label{conserv-stress-2}
\end{equation}
where we have introduced the initial value of the density $\rho_0\equiv \rho(0).$

Substituting~\eqref{conserv-stress-2} into~\eqref{eomin-00}, evaluating the resulting equation at $\tau=0$ and imposing the initial condition $\dot{a}(0)=0$ (the dust configuration starts at rest), we obtain the following relation between the spatial curvature $k,$ the initial density $\rho_0$ and the initial acceleration $\ddot{a}(0):$
\begin{equation}
\rho_0=\frac{3}{8\pi G_{\rm N}}k+3\alpha\left(k^2-\ddot{a}^2(0)\right)\,.
    \label{paremeters-relation}
\end{equation}
If we set $\alpha=0$ we consistently recover the known relation between $k$ and $\rho_0$ that is valid in the case of the Oppenheimer-Snyder collapse in GR~\cite{Oppenheimer:1939ue}. 

\subsubsection{Field equations in dimensionless form}

It is convenient to rewrite all the relevant equations in dimensionless form. This can be done by rescaling the dimensionful quantities by the mass scale $m_0=(8\pi G_{\rm N}\alpha)^{-1/2}$. Doing so, eqs.~\eqref{eomin-00} and~\eqref{paremeters-relation} become
\begin{eqnarray}
3\frac{1}{a^2}\left(\tilde{k}+{a^\prime}^2\right) +3\left[ \frac{1}{a^4}\left(\tilde{k}^2-2 \tilde{k} {a^\prime}^2-3{a^\prime}^4 \right)-\frac{{a^{\prime\prime}}^2}{a^2} +2\frac{{a^\prime}a^{\prime\prime\prime}}{a^2}+2\frac{{a^\prime}^2a^{\prime\prime}}{a^3}\right] =\frac{\tilde{\rho}_0}{a^3}\,, \label{eomin-00-dimensionless}
\end{eqnarray}
and
\begin{equation}
\tilde{\rho}_0=3\tilde{k}+3\left(\tilde{k}^2-{a^{\prime\prime}}^2(0)\right)\,,
    \label{paremeters-relation-dimensionless}
\end{equation}
respectively, where we have defined the dimensionless quantities 
\begin{equation}
\tilde{\rho}_0\equiv \frac{8\pi G_{\rm N}}{m_0^2}\rho_0\,,\qquad \tilde{k}\equiv \frac{k}{m_0^2}\,,\qquad \tilde{\tau}\equiv m_0\tau\,,\qquad ^\prime\equiv \frac{{\rm d}}{{\rm d}\tilde{\tau}}\,.
\label{dimensionless-quantities}
\end{equation}
We can also introduce the dimensionless version of the radial coordinate, i.e. $\tilde{\chi}  \equiv m_0\chi,$ and thus define the dimensionless radius of the star to be $\tilde{\chi}_*=m_0 \chi_*.$ With these definitions we now measure distances and time scales in unit of $1/m_0.$

In the case of Einstein's GR the $(\tau\tau)$-component is a first-order differential equation, i.e.  it is not dynamical, thus we have to solve the trace equation that, in dimensionless form, reads\footnote{A comment is in order. Obviously in the case of GR we do not have any contribution proportional to $1/m_0$ in the field equations. In eqs.~\eqref{trace-int-dimensionless-GR} and~\eqref{continuity-dimensionless-gr} we have used $1/m_0$ simply as an artificial physical length scale to rewrite the equations in dimensionless form and in order to make a consistent comparison between the solution in GR and the one in Quadratic Gravity.}
\begin{eqnarray}
6\left[ \frac{1}{a_{\rm GR}^2}\left(\tilde{k}_{\rm GR}+a_{\rm GR}^{\prime \,2}\right)+\frac{a_{\rm GR}^{\prime\prime}}{a_{\rm GR}} \right]=-\frac{\tilde{\rho}_0}{a_{\rm GR}^3}\,,
\label{trace-int-dimensionless-GR}
\end{eqnarray}
where we have introduced the GR analogue of the scale factor, $a_{\rm GR},$ and the rescaled spatial curvature, $\tilde{k}_{\rm GR},$ that can differ from the Quadratic Gravity value $\tilde{k}$ and satisfies the relation 
\begin{eqnarray}
\tilde{\rho}_0=3\tilde{k}_{\rm GR}\,.
\label{continuity-dimensionless-gr}
\end{eqnarray}

\subsection{Interior solution}

The interior field equations cannot be solved analytically but we can solve them numerically. Since there is only one unknown function, i.e. the scale factor $a(\tau),$ it is sufficient to focus on the $(\tau\tau)$-component in~\eqref{eomin-00} which contains up to third-order derivatives\footnote{If instead we wanted to focus on the trace equation~\eqref{trace-int}, then we would solve a fourth-order differential equation that requires an initial condition for the third-order derivative. This additional initial condition is not arbitrary but is fixed by the solution of the $(\tau\tau)$-component in~\eqref{eomin-00}.}.

\subsubsection{Initial conditions}

Since eq.~\eqref{eomin-00-dimensionless} is a third-order differential equation, we need to impose three initial conditions: $a(0),$ $a^\prime(0)$ and $a^{\prime\prime}(0)$.  As already mentioned above, we set $a(0)=1$ and assume that the uniform ball of dust starts at rest, i.e. $a^\prime(0)=0.$ As for the value of $a^{\prime\prime}(0)$, in principle, we could consider different scenarios. 

The physically interesting configuration is that in which the initial phase of the collapse is not affected by new physics beyond GR.
Indeed, an astrophysical collapse typically starts in regimes of low energy density where standard known physics should still apply. Higher curvature effects are expected to become important later in the collapse, when the star has sufficiently contracted, so that $\alpha R^2$ can dominate over $\frac{1}{G}R.$

Therefore, to describe a physically valid scenario we have to consider initial conditions in which GR still holds. We choose the initial value of the acceleration to be equal to that of GR, i.e. $a^{\prime\prime}(0)=a_{\rm GR}^{\prime\prime}(0)=-\tilde{k}_{\rm GR}/2=-\tilde{\rho}_0/6$. 

\subsubsection{Numerical solution}

To solve eq.~\eqref{eomin-00-dimensionless} we expressed $\tilde{k}$ as a function of $\tilde{\rho}_0$ through the relation 
\begin{equation}
\tilde{k}=\frac{1}{2}\left[ \sqrt{1+\frac{4}{3}\left(\tilde{\rho}_0+\frac{\tilde{\rho}_0^2}{12}\right)}-1 \right]\,,
\end{equation}
so that the only remaining parameter to specify in order to determine the solution is $\tilde{\rho}_0.$

\begin{figure}[t!]
	\centering
\subfloat[Subfigure 1 list of figures text][]{
		\includegraphics[width=0.47\linewidth]{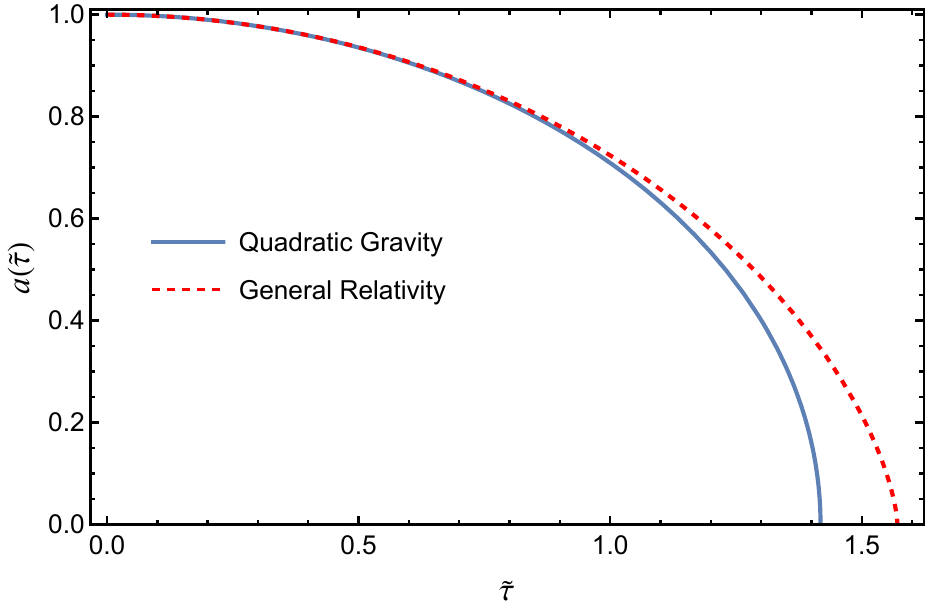}\label{fig1.1}}\quad\,
	\subfloat[Subfigure 2 list of figures text][]{
		\includegraphics[width=0.47\linewidth]{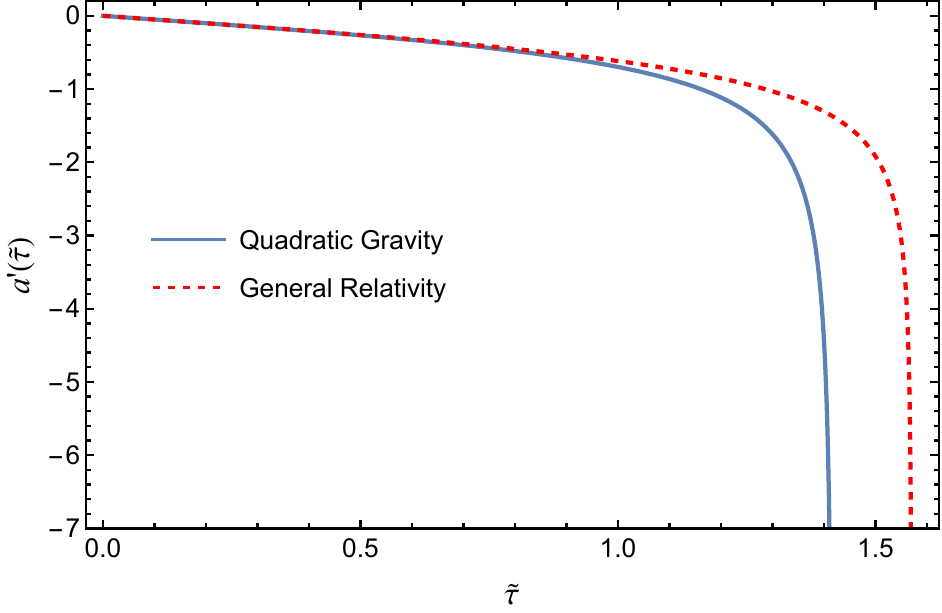}\label{fig1.2}}\\
  \subfloat[Subfigure 2 list of figures text][]{
		\includegraphics[width=0.47\linewidth]{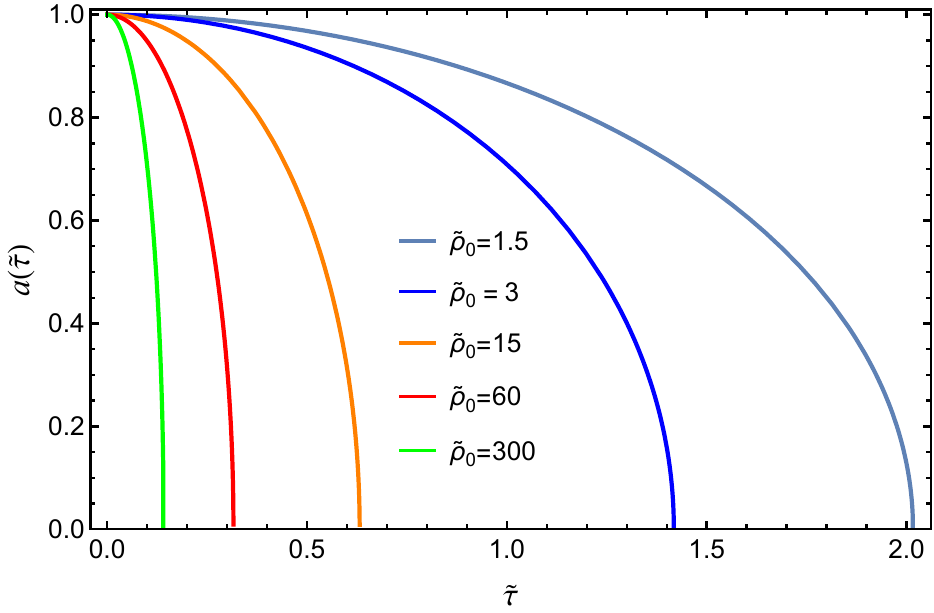}\label{fig1.4}}\quad\,
  \subfloat[Subfigure 2 list of figures text][]{
		\includegraphics[width=0.47\linewidth]{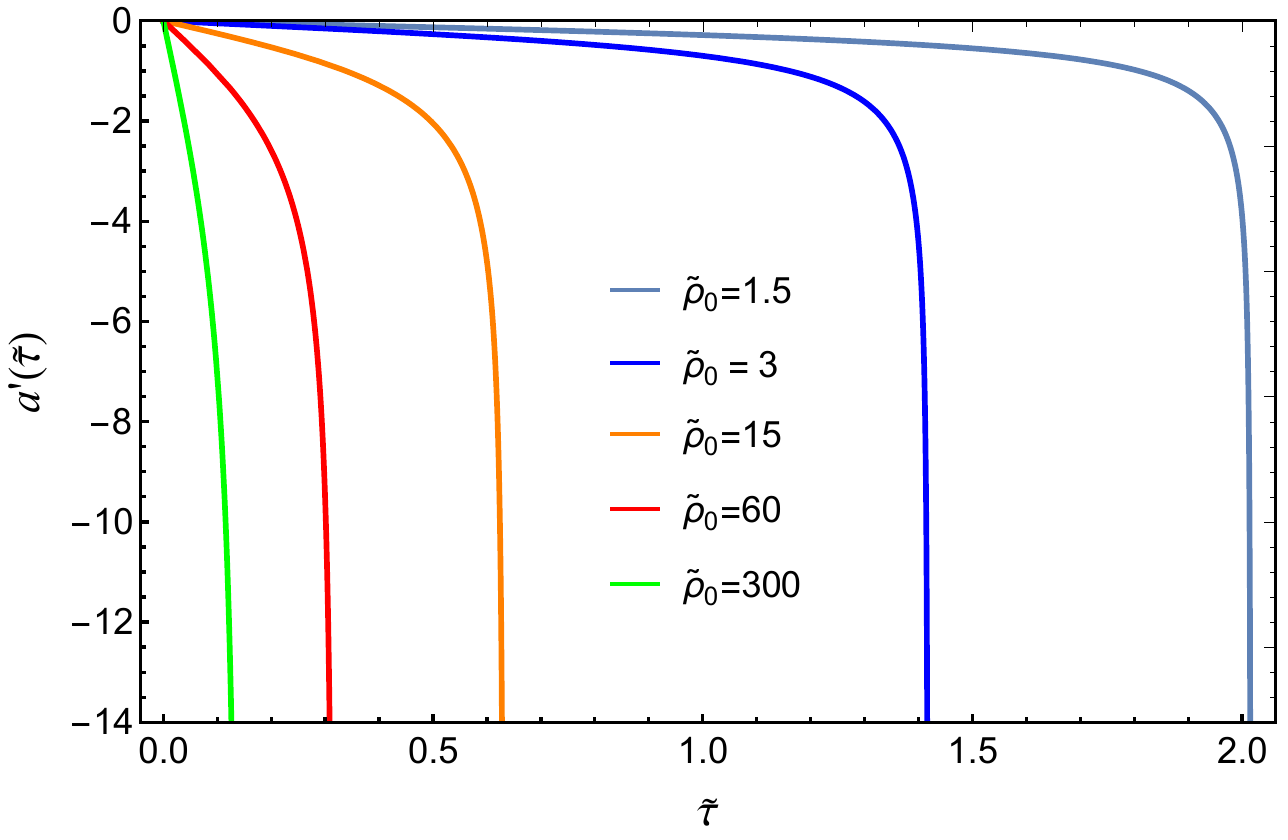}\label{fig1.5}}
 	\protect\caption{(a)~Numerical solution of the Quadratic Gravity field equation~\eqref{eomin-00-dimensionless} (purple line) in comparison with the Oppenheimer-Snyder in GR that is found by solving~\eqref{trace-int-dimensionless-GR}  (red dotted line). We set $\tilde{\rho}_0=3.$ (b)~corresponds to the analogue plots for the velocity $a^\prime (\tilde{\tau})$.
   (c)~Numerical solutions for the scale factor and (d)~its derivative in Quadratic gravity in the cases $\tilde{\rho}_0=1.5$ (purple line), $\tilde{\rho}_0=3$ (blue line), $\tilde{\rho}_0=15$ (orange line), $\tilde{\rho}_0=60$ (red line) and $\tilde{\rho}_0=300$ (green line).}
  \label{fig1}
\end{figure}

The numerical solution of eq.~\eqref{eomin-00-dimensionless} is shown in Fig.~\ref{fig1}. 
In particular, in Fig.~\ref{fig1.1} we compared the Quadratic Gravity solution to the case of Oppenheimer-Snyder in GR (i.e. with $\alpha=0$). We can clearly notice that in Quadratic Gravity the collapse still happens and is faster. In Figs.~\ref{fig1.2}  we plotted the first derivative of the scale factor. Both the velocity and the acceleration diverge negatively at the end of the collapse. 
In Fig.~\ref{fig1.4} and Fig.~\ref{fig1.5} we plotted the Quadratic Gravity numerical solution for different values of the initial density. We found that for larger values of $\tilde{\rho}_0$ the collapse is faster, still in agreement with the physical expectation. 

\paragraph{Remark.} It is worth mentioning that to make the deviations from GR during the collapse more appreciable, we have chosen values of $\rho_0$, and thus of $k$ and $k_{\rm GR},$ such that the initial radius of the star $\chi_*$ is comparable to or smaller than $1/m_0$. Indeed, since we chose values of $\tilde{k}_{\rm GR}$ comparable or larger than $\mathcal{O}(1)$ and given the condition $\tilde{\chi}_*<1/\sqrt{\tilde{k}_{\rm GR}},$ it follows that $\chi_*$ is comparable or smaller than $1/m_0.$ However, in typical scenarios we expect $\chi_*\gg 1/m_0.$ We studied this case too and obtained the same qualitative result. The values of the parameters in the plots were chosen to make the deviations from GR more evident.


\subsubsection{Alternative initial conditions}

So far we have solved the field equations by making the physically motivated choice that the initial acceleration is equal to the GR value, i.e. $a^{\prime\prime}(0)=a^{\prime\prime}_{\rm GR}(0).$ This was enough to ensure that at the beginning of the collapse well-known physics still applies and that new effects due to the higher curvature terms become important at later times when the matter has contracted sufficiently. However, we can always ask what happens to the evolution of the scale factor if we assume different initial conditions for the acceleration. In fact, we could also consider the following two alternative scenarios:
\begin{itemize}

\item We can assume that the initial acceleration is different from the GR value but the spatial curvatures are the same, i.e. $\tilde{k}=\tilde{k}_{\rm GR}.$ Then, from eq.~\eqref{paremeters-relation-dimensionless} we obtain $a^{\prime\prime}(0)=-\tilde{k}_{\rm GR}$ and the relation~\eqref{paremeters-relation} between the initial density and the spatial curvature does not depend on the value of the quadratic-curvature coefficient $\alpha.$

\item We can choose $a^{\prime\prime}(0)$ to have an arbitrary small and negative value. This also includes scenarios where higher-curvature effects may be dominant already in the initial phase of the matter contraction.

\end{itemize}

In the first case, we checked that the collapse still happens and the behavior of the scale factor is similar to the one found above when the initial acceleration is equal to the GR value. In fact, this scenario can still be considered as physically viable because the initial conditions are very similar to those in GR since $k=k_{\rm GR}=-a^{\prime\prime}(0)$ implies $a^{\prime\prime}(0)=2a^{\prime\prime}_{\rm GR}(0),$ which means that the two initial accelerations are still of the same order of magnitude.

In the second case, in principle anything can happen. However, if $|a^{\prime\prime}(0)|$ is much bigger or smaller than $|a^{\prime\prime}_{\rm GR}(0)|= \tilde{k}_{\rm GR}/2,$ we are in a regime in which the higher-curvature terms are important already during the initial phase of the collapse. In fact, in such a case the field equations at $\tau=0$ are satisfied if and only if some cancellation between linear and quadratic terms happens. This situation is unphysical because at the initial stage of an astrophysical collapse we expect GR to hold and higher-curvature effects to be negligible. On the other hand, if we choose values of the initial acceleration $a^{\prime\prime}(0)$ that are not too different from $a^{\prime\prime}_{\rm GR}(0)$ then the configuration could still be considered physically viable and we have checked that a collapse always happens.

To understand the main results of the next section it is sufficient to only focus on the solution compatible with the initial conditions according to which the initial acceleration is equal to the GR value: $a^{\prime\prime}(0)=a^{\prime\prime}_{\rm GR}(0)=-\tilde{k}_{\rm GR}/2$.

\subsubsection{Analytic late-time solution}
Finally, we have studied the behavior of the scale factor near the formation of the singularity. To this end, we simply make the ansatz 
\begin{equation}
    a(\tau)\propto \left(\tau_0-\tau\right)^\gamma\,,
\end{equation}
that is valid in the regime $\tau\sim \tau_0,$ where $\tau_0$ marks the proper time of the end of the collapse. In the GR case, truncating the field equation at the leading order we get $\gamma=2/3$. In contrast to that, Quadratic Gravity yields $\gamma=1/2$. We have also checked numerically that this behavior provides a good approximation for the late-time evolution. 
This confirms that, as shown by the plot in Fig.~\ref{fig1}, the final stage of collapse is faster than its counterpart in GR.

\section{Horizon formation}\label{sec:horiz-form}

We can now ask whether the solution we have found admits the formation of a horizon  at some point during the collapse. Before doing this, it might be useful to specify which notion of horizon we are referring to.

Black holes are sometimes defined as the spacetime regions within event horizons. An event horizon is a causal boundary within which signals cannot reach the asymptotic region. However, this is a global notion and no local observer can probe the presence of the event horizon \cite{Hayward:1993mw,Visser:2014zqa}. We, therefore, study the presence of a trapping horizon which is a (quasi)local entity and it is defined as the boundary of the spacetime region that is locally trapped. Finally, it is useful to consider the apparent horizon that is a 2-dimensional section of the trapping horizon.\footnote{Note that the notion of apparent horizon is foliation dependent. While in spherical symmetry we do not need to deal with this ambiguity, in more generic spacetimes we might prefer to refer to the notion of dynamical horizons. We refer readers interested in these issues and in formal definitions of different notions of horizons to reviews on the field~\cite{Ashtekar:2003hk,Ashtekar:2004cn}.}

\begin{figure}[t!]
	\centering
\subfloat[Subfigure 1 list of figures text][]{
		\includegraphics[height=6.5cm]{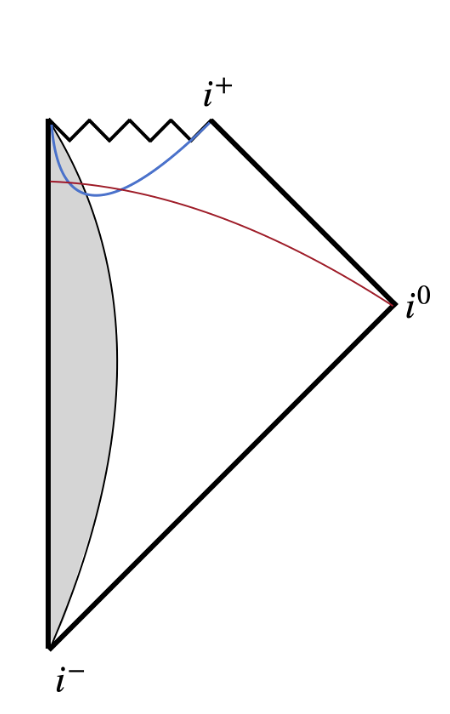}\label{fig3.1}}\hspace{3cm}\,
	\subfloat[Subfigure 2 list of figures text][]{
		\includegraphics[height=6cm]{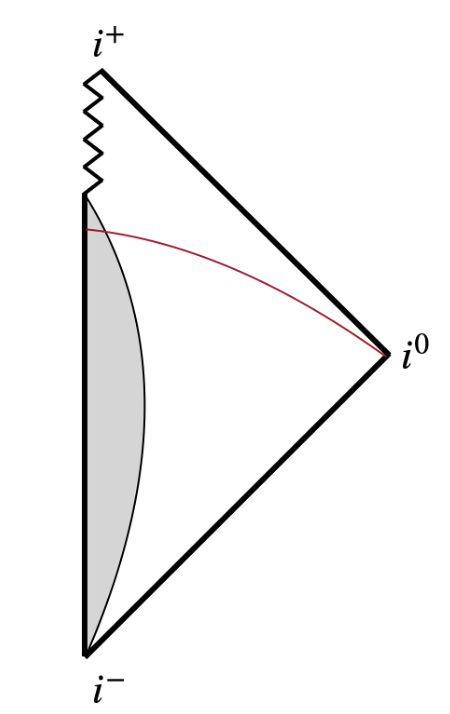}\label{fig3.2}}
 	\protect\caption{Conformal diagrams of possible spacetimes describing gravitational collapse into a singularity. (a)~The left diagram shows a spacetime with a trapping horizon (blue line). Each Cauchy hypersurface (red line) crosses the trapping horizon an even number of times.  (b)~The right diagram shows a collapse into a naked singularity. This second possibility is excluded in our analysis due to the presence of an inner horizon in the interior region.}
  \label{fig3}
\end{figure}

The presence of a trapping horizon can be investigated by focusing solely on the internal geometry. 
In fact, if a trapped surface (i.e. an inner apparent horizon) forms in the interior region, then an outer apparent horizon must also be present in order to recover the correct light-cone structure in the asymptotic region. The situation is illustrated in Fig.~\ref{fig3}. The shaded area represents the portion of spacetime inside the collapsing matter. 
We can study the casual structure in this region. 
If we find a trapped surface and an inner apparent horizon (Fig.~\ref{fig3.1}), we know that an outer apparent horizon must be present in the exterior region. On the contrary, if there is no inner horizon, it is possible that the spacetime describes a naked singularity (Fig.~\ref{fig3.2}).

Trapped regions can be found by looking at the expansion parameters of the two vectors spanning a null geodesic congruence. For the metric in eq.~\eqref{interior-ansatz} the outgoing and ingoing future directed null vectors are 
\begin{equation}
    \boldsymbol{\ell}=\frac{1}{\sqrt{2}}\partial_\tau+\frac{1}{\sqrt{2}}\frac{\sqrt{1-k\chi^2}}{a(\tau)}\partial_\chi\,,\qquad
    \boldsymbol{k}=\frac{1}{\sqrt{2}}\partial_\tau-\frac{1}{\sqrt{2}}\frac{\sqrt{1-k\chi^2}}{a(\tau)}\partial_\chi\,,
\end{equation}
respectively. It is easy to show that the following relations are satisfied: $\ell_\mu \ell^\mu =0,$ $k_\mu k^\mu =0,$ $\ell_\mu k^\mu =-1$. The null vectors are defined up to an irrelevant overall factor. 

Introducing the tensor
\begin{equation}
q_{\mu\nu}=g^-_{\mu\nu}+k_\mu \ell_\nu + k_\nu \ell_\mu
\end{equation}
that projects onto the two-dimensional subspace orthogonal to $k^\mu$ and $\ell^\mu,$ where $g^{-}_{\mu\nu}$ are the components of the interior metric in eq.~\eqref{interior-ansatz}, we can define the two expansion parameters as 
\begin{equation}
\theta^{(\boldsymbol{\ell})}=q^{\mu\nu}\nabla_\mu\ell_\nu\,,\qquad \theta^{(\boldsymbol{k})}=q^{\mu\nu}\nabla_\mu k_\nu\,.
\end{equation}
Their explicit expressions in terms of the scale factor are 
\begin{equation}
   \theta^{(\boldsymbol{\ell})}= \sqrt{2}\frac{1-k\chi^2+ \chi \sqrt{1-k\chi^2} \dot{a}(\tau)}{\chi a(\tau) \sqrt{1-k \chi^2}}\,,\qquad
   \theta^{(\boldsymbol{k})}= -\sqrt{2}\frac{1-k\chi^2 - \chi \sqrt{1-k\chi^2} \dot{a}(\tau)}{\chi a(\tau) \sqrt{1-k \chi^2}}\,.
\end{equation}

The expansions encode information about how the area of the two-dimensional cross-section of a null congruence varies in the outgoing and ingoing directions. In the flat space limit we get $\theta^{(\boldsymbol{\ell})}=\sqrt{2}/\chi$ and $\theta^{(\boldsymbol{k})}=-\sqrt{2}/\chi,$ that is, in Minkowski spacetime the expansion of ingoing light rays is always negative ($\theta^{(\boldsymbol{k})}<0$) and that of outgoing light rays is always positive ($\theta^{(\boldsymbol{\ell})}>0$). In our context, to show that a trapped surface is formed, we need to verify that both expansion parameters can become negative at some point during the collapse. In particular, the condition $\theta^{(\boldsymbol{\ell})}=0$ corresponds to the formation of a marginal trapped surface, i.e., of an apparent horizon.

Since $\dot{a}$ is negative and diverges at the end of the collapse ($\dot{a}\rightarrow -\infty$), it follows that $\theta^{(\boldsymbol{k})}$ is always negative as the term $\chi \sqrt{1-k\chi^2}\dot{a}(\tau)$ in the numerator dominates. Whereas, $\theta^{(\boldsymbol{\ell})}$ starts with a positive value but becomes negative at some point during the collapse, in particular it is zero when
\begin{equation}
\theta^{(\boldsymbol{\ell})}(\tau)=0\quad \Leftrightarrow \quad \chi(\tau)=\frac{1}{\sqrt{k+\dot{a}^2(\tau)}}\equiv \chi_{\rm ah}(\tau)\,,
\end{equation}
where $\chi_{\rm ah}(\tau)$ is defined to be the radius of the apparent inner horizon. Its initial value $\chi_{\rm ah}(0)=1/\sqrt{k}$ is located outside the star since $1/\sqrt{k}>\chi_*$, therefore it is unphysical as the coordinate $\chi$ is only defined inside the star, i.e. $\chi\leq \chi_*.$ We can say that the apparent horizon forms at the time $\tau_{\rm ah}$ when $\chi_{\rm ah}(\tau_{\rm ah})=\chi_*.$ Since the derivative of the scale factor tends to infinity at some finite time $\tau>\tau_{\rm ah},$ it follows that the inner apparent horizon goes to zero and ends up into the singularity.

\begin{figure}[t!]
	\centering
\subfloat[Subfigure 1 list of figures text][]{
		\includegraphics[scale=0.46]{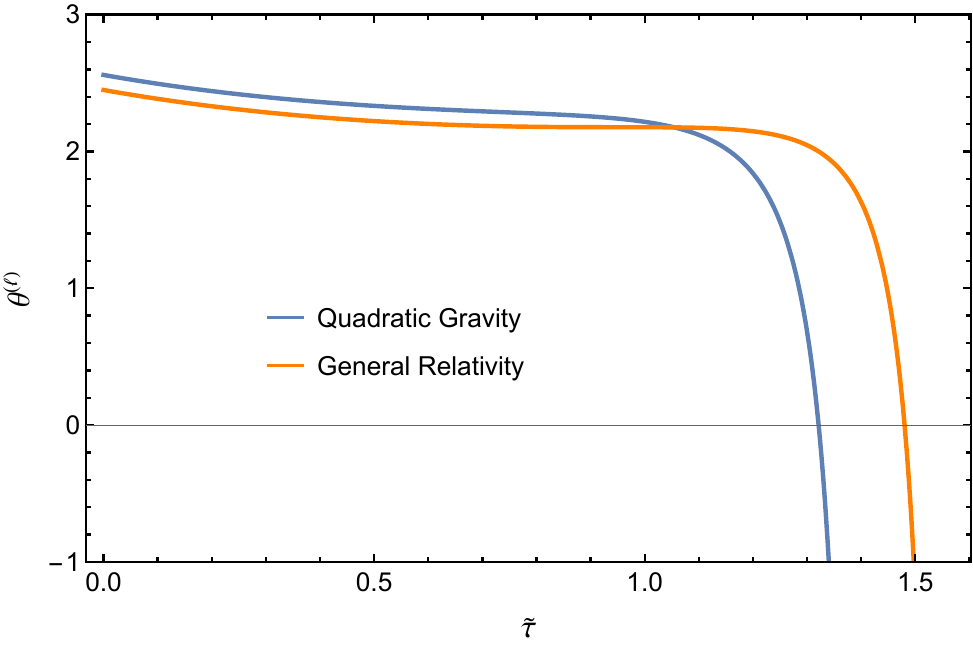}\label{fig2.1}}\quad\,
	\subfloat[Subfigure 2 list of figures text][]{
		\includegraphics[scale=0.47]{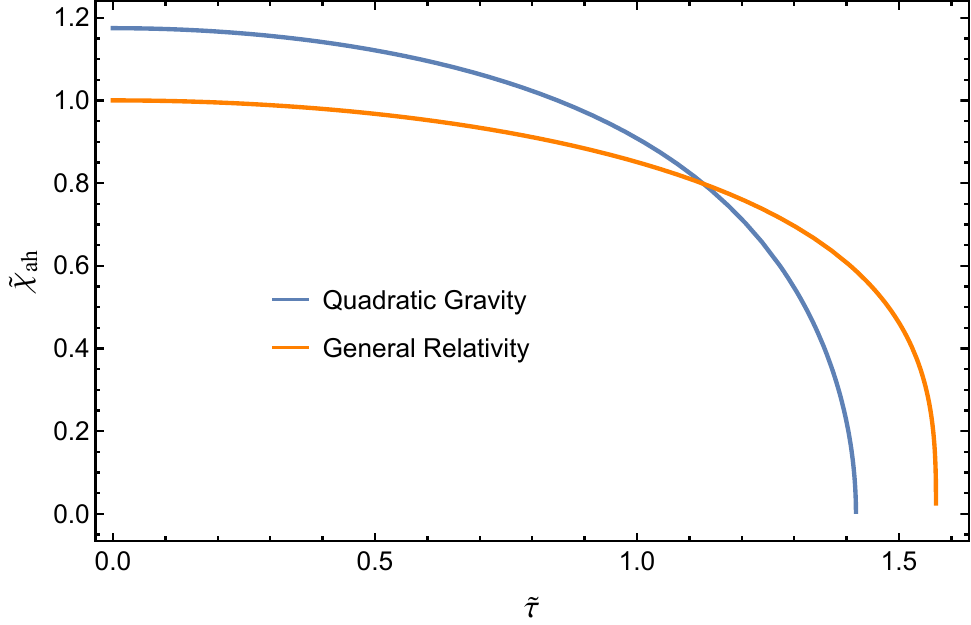}\label{fig2.2}}
 	\protect\caption{(a) Behavior of the outgoing expansion $\theta^{(\boldsymbol{\ell})}$ in both cases of Quadratic Gravity (purple line) and Einstein's GR (orange line). We set $\tilde{\chi}_*\equiv m_0\chi_*=0.5$. (b) Behavior of the apparent horizon curve $\tilde{\chi}_{\rm ah}(\tilde{\tau})$ (i.e. the surface at which $\theta^{(\boldsymbol{\ell})}=0$) in both cases of Quadratic Gravity (purple line) and Einstein's GR (orange line). We set $\tilde{\rho}_0=3$ in both plots.}
  \label{fig2}
\end{figure}

The plot in Fig.~\ref{fig2.1} shows the behavior of the outgoing expansion $\theta^{(\boldsymbol{\ell})}$ as a function of the comoving time $\tau$ and for $\tilde{\chi}_*=0.5$. The expansion becomes negative at the time scale $\tau_{\rm ah}$ when $\tilde{\chi}_{\rm ah}(\tau_{\rm ah})=\tilde{\chi}_*=0.5$. In Fig.~\ref{fig2.2} we have shown the behaviour of the apparent horizon $\chi_{\rm ah}(\tau)$ which goes to zero at the singularity, i.e. when $\dot{a}\rightarrow -\infty.$ Note that the curves in Figs.~\ref{fig2.1} and~\ref{fig2.2} intersect because $k\neq k_{\rm GR}.$ This fact implies that, depending on the value of $\chi_*$, in Quadratic Gravity the trapped region can form earlier or later than in the case of GR. It is also worth to mention that if we choose the initial condition for the acceleration $\ddot{a}(0)=-k$ which implies $k=k_{\rm GR},$ then the horizon in Quadratic Gravity would form always earlier than in GR.


\paragraph{Main result.} The formation of an inner horizon also implies the presence of an outer horizon as illustrated in Fig.~\ref{fig3.1}. As a consequence, also the exterior spacetime metric must possess a horizon in order to be compatible with the interior solution and to recover the correct light-cone structure in the asymptotic region.
This excludes 2-2 holes~\cite{Holdom:2016nek,Holdom:2019bdv} and other horizonless ultra-compact objects~\cite{Salvio:2019llz} as possible solutions for the exterior metric that can match a collapsing star with uniform dust density and homogeneous and isotropic interior geometry. In particular, spacetimes with a naked singularity, whose conformal diagram is illustrated in Fig.~\ref{fig3.2}, are  not allowed.  Therefore, black hole metrics are the ones that could possibly be matched to the interior solution found above.

\paragraph{Remark.} For the sake of completeness, we also checked that if we choose alternative initial conditions for which higher curvature effects are relevant already at the beginning of the collapse, then it would be possible to find interior solutions that do not form a horizon. However, as already explained above, these solutions would correspond to unphysical astrophysical scenarios since we expect well-known physics to hold at the beginning of a collapse.

\section{Towards constraining the exterior metric}\label{sec:exter}

The next question to ask is what exterior spacetime metric can be smoothly matched to the interior solution. Finding an exact answer to this question is very difficult because we have to solve the full set of differential equations in the outer region for the spherically symmetric metric ansatz
\begin{equation}
({\rm d}s^2)_+=g^+_{\mu\nu}(x_+){\rm d}x_+^\mu{\rm d}x_+^\nu=-A(t,r){\rm d}t^2+B(t,r){\rm d}r^2+r^2({\rm d}\theta^2+\sin^2\theta {\rm d}\varphi^2)\,,
\label{ext-ansatz-t-r}
\end{equation}
where $x_+^{\mu}=(t,r,\theta,\varphi)$ are the exterior coordinates, and find the solutions for the two unknown functions $A(t,r)$ and $B(t,r).$ The complexity is mainly related to three points: (i) the presence of higher-order curvatures and higher-order derivatives make the structure of the field equations and the junction conditions much more complicated; (ii) the metric functions depend on both time and space which lead to a system of non-linear partial differential equations; (iii) in Quadratic Gravity there is no Birkhoff's theorem that singles out one vacuum spherically symmetric solution.

In this paper we will not find an exact exterior solution. However, we will introduce all the ingredients that are needed to tackle the problem and we will use various arguments to partly constrain the form of the metric functions $A(t,r)$ and $B(t,r).$ Our analysis will set the stage for future investigations aimed at fully determining the spacetime metric outside a spherically symmetric collapsing star in the context of Quadratic Gravity.

In what follows we will first derive the generalized junction conditions in the case of Quadratic Gravity. Subsequently, we will prove several no-go theorems through which we can constrain some features of the exterior metric. In particular, we will learn that the spacetime metric that can be possibly matched to the collapsing uniform-density dust star with homogeneous and isotropic interior geometry cannot be stationary. This means that stationarity (i.e. staticity in the spherically symmetric case) may only be reached asymptotically at late times and for distances larger than the star's radius. Finally, we will provide some arguments to further constrain the exterior metric.

\subsection{Useful geometric quantities}

Let us introduce several ingredients that are needed to write down the junction conditions; a similar treatment was followed in Ref.~\cite{Casado-Turrion:2022xkl} in the case of $f(R)$ theories. 

The surface of the star on the outer side is described in terms of the coordinates $x^\mu_{+*} =(t_*,r_*,\theta,\varphi),$ where 
\begin{equation}
t_*=t_*(\tau)\,,\qquad r_*=r_*(\tau)\,,
\end{equation}
are the exterior time and radial coordinates evaluated at the star's surface. The angular coordinates can be chosen to be the same both on the inner and outer sides due to spherical symmetry. Furthermore, we can define the coordinates on the three-dimensional surface of the star to be $y^a=(\tau,\theta,\varphi).$

We can characterize the  star's surface in terms of normal and tangent vectors defined on the inner and outer sides. The tangent vectors on the inner and outer sides are respectively given by
\begin{equation}
e^{-\mu}_a\equiv \frac{\partial x^\mu_{-*}}{\partial y^a}=\delta^\mu_a\,,\qquad e^{+\mu}_a\equiv \frac{\partial x_{+*}^\mu}{\partial y^a}= \left( \dot{t}_*\delta_t^\mu+\dot{r}_*\delta^\mu_r\right)\delta_a^\tau + \delta_\theta^\mu \delta_a^\theta + \delta_\varphi^\mu \delta_a^\varphi\,.
\end{equation}
The normal vectors $n^\pm_\mu$ on the two sides can be found by imposing the conditions $n^\pm_\mu n^{\pm\mu}=1$ and $n^\pm_\mu e^{\pm\mu}_a=0$, and we get
\begin{equation}
n^-_\mu =\frac{a(\tau)}{\sqrt{1-k\chi_*^2}}\delta_\mu^\chi\,,\qquad n^+_\mu =\frac{\sqrt{A_* B_*}}{\sqrt{\dot{t}_*^2A_* - \dot{r}_*^2}B_*}\left(-\dot{r}_* \delta_\mu^t+\dot{t}_* \delta_\mu^r\right)\,,
\end{equation}
where we have defined $A_*\equiv A(t_*(\tau),r_*(\tau))$ and $B_*\equiv B(t_*(\tau),r_*(\tau))$.

We can also define the projector 
\begin{equation}
h^{\pm}_{\mu\nu}=g^{\pm}_{\mu\nu}-n^{\pm}_\mu n^\pm_\nu\,,
\end{equation}
that is tangent to the star's surface, i.e. $n^{\pm\mu} h^{\pm}_{\mu\nu}=0$.

Using the tangent vectors we can define the three-dimensional metrics $h^\pm_{ab}$ induced on both sides of star's surface as
\begin{equation}
({\rm d}s^2)_{\pm *}\equiv  h^\pm_{ab}{\rm d}y^a{\rm d}y^b\,,\qquad 
h^{\pm}_{ab}=e^{\pm\mu}_a e^{\pm\nu}_b g^{\pm}_{\mu\nu}=e^{\pm\mu}_a e^{\pm\nu}_b h^{\pm}_{\mu\nu}\,.
\end{equation}
The explicit expressions for the induced metrics are
\begin{eqnarray}
h^-_{ab}&=& \text{diag}\left(-1,\, a(\tau)\chi_*^2,\, a(\tau)\chi_*^2 \sin^2\theta \right)\,, \\[1.5mm]
h^+_{ab}&=& \text{diag}\left(-(\dot{t}_*^2 A_*-\dot{r}_*^2 B_*),\, r_*^2,\, r_*^2 \sin^2\theta \right)\,.
\end{eqnarray}

We can do the same for the extrinsic curvature and we get
\begin{equation}
K^\pm_{ab}=-n_\mu^\pm \left( \frac{\partial e^{\pm \mu}_b}{\partial y^a}+\Gamma^\mu_{\pm \rho\sigma} e^{\pm \rho}_a e^{\pm \sigma}_b \right)\,.
\end{equation}
The non-vanishing components on the inner side are
\begin{equation}
K^-_{\theta\theta}=a(\tau)\chi_*\sqrt{1-k\chi_*^2}=\frac{1}{\sin^2\theta}K^-_{\varphi\varphi}\,,
\end{equation}
while the expressions of those on the outer side are\footnote{For the sake of clarity, let us mention that whenever we use partial derivatives together with the evaluation at the star's surface, the former is always performed first, e.g., $\partial_t A_*$ means $(\partial_t A)|_*.$} 
\begin{eqnarray}
K^+_{\tau\tau}&=& \frac{-1}{2\sqrt{A_*B_*}\sqrt{A_*\dot{t}^2_*-B_*\dot{r}^2_*}} \left[A_* \dot{t}_*\left(\dot{t}^2_*\partial_rA_*+2\dot{r}_*\dot{t}_*\partial_t B_*+\dot{r}^2_*\partial_r B_*\right) \right.\nonumber \\[1.5mm] 
&&\left.- B_* \dot{r}_*\left(\dot{t}^2_*\partial_t A_*+2\dot{r}_*\dot{t}_*\partial_r A_*+\dot{r}^2_*\partial_t B_*\right)+2A_*B_*\left(\dot{t}^2_*\ddot{r}_*-\dot{r}^2_*\ddot{t}_*\right)   \right]\,,\\[2mm]
K^+_{\theta\theta}&=& \sqrt{\frac{A_*}{B_*}}\frac{r_*\dot{t}_*}{\sqrt{\dot{t}^2_*A_*-\dot{r}^2_*B_*}} = \frac{K^+_{\varphi\varphi}}{\sin^2\theta}\,.
\end{eqnarray}

\subsection{Junction conditions}

The general form of the junction conditions in Quadratic Gravity for the action~\eqref{action-weyl->Ricci} with $\alpha\neq 0$ and $\beta\neq 0$ were first derived in Ref.~\cite{Reina:2015gxa} and are given by
\begin{subequations}
    \begin{align}
\left[ h_{ab} \right] &= 0\,, \qquad \left[ K_{ab} \right] = 0 \label{1-2-junction}\\[1.5mm]
\left[ R_{\mu\nu} \right] &= 0\,, \qquad \left[ \nabla_\rho R_{\mu\nu} \right] = 0 \label{3-4-junction}
\end{align}
\end{subequations}
where the square-brackets notation is defined as $[F]\equiv F^+-F^-$ for any quantity $F.$ 
In addition to the well-known first and second Darmois-Israel junction conditions~\cite{Darmois1927,Israel:1966rt} in eq.~\eqref{1-2-junction} we also have two additional conditions in eq.~\eqref{3-4-junction} that we call \textit{third} and \textit{fourth} junction conditions, respectively. 

In our case we are interested in the so-called proper matching conditions for which no distributional matter is present at the junction, i.e. there is no discontinuity at the star's surface. We now want to determine the independent number of components of the junction conditions and study how they constrain the exterior metric solution.

\begin{itemize}

\item \textbf{First junction condition:}  It gives two independent conditions:
\begin{equation}
r_*=a(\tau)\chi_*\,,\qquad \dot{t}_*^2A_* - \dot{r}_*^2 B_* =1\,. \label{1st-junc-cond}
\end{equation}
The condition $[h_{ab}]=0$ also implies that on the surface (i.e. for $\chi=\chi_*$, $t=t_*$ and $r=r_*$) we have $[h_{\mu\nu}]=0$ and $[n_\mu]=0.$

\item \textbf{Second junction condition:} It gives two conditions that in general are independent. Using the first junction, the conditions coming from the second junction can be written as
\begin{equation}
\beta=\beta_0\sqrt{A_*B_*}\,,\qquad \dot{\beta}=\frac{1}{2}\left( \dot{t}^2_*\partial_t A_*  - \dot{r}^2_* \partial_t B_*  \right)\,, 
\label{2nd-junc-cond}
\end{equation}
where we have defined the quantities $\beta\equiv A_* \dot{t}_*$ and $\beta_0\equiv \sqrt{1-k\chi_*^2}$. Note that in the case of the Schwarzschild metric in GR we have $A_*B_*=1$ and $\dot{\beta}=0,$ thus only one independent condition. However, in the more general case with $B_*\neq 1/A_*$ we have two independent conditions.

\item \textbf{Third junction condition:} Using $[h_{ab}]=0$ and $[K_{ab}]=0,$ the third junction condition can be written as~\cite{Reina:2015gxa}
\begin{equation}
0=[R_{\mu\nu}]=h^\rho_\mu h^\sigma_\nu [R_{\rho\sigma}]+\frac{1}{2}n_\mu n_\nu [R]\,;
\end{equation}
which is equivalent to the two conditions
\begin{equation}
\qquad [R]=0\,,\qquad e^\mu_a e^\nu_b [R_{\mu\nu}]=0\,.
\end{equation}
Note that the tangent-normal component of $[R_{\mu\nu}]$ does not contribute when $[K_{ab}]=0.$
The first equation is the continuity of the Ricci scalar 
\begin{equation}
R^+_*= R^-_*= 6\left( \frac{\dot{a}^2+k}{a^2}+\frac{\ddot{a}}{a}\right) \,,\label{3rd-junc-cond-normal-normal}
\end{equation}
while the second gives two independent conditions
\begin{eqnarray}
\dot{t}_*^2 R^+_{tt*}+2\dot{t}_* \dot{r}_* R^+_{tr*}+\dot{r}^2_* R^+_{rr*}&=&-3\frac{\ddot{a}}{a}\,, \\[1.5mm]
R^+_{\theta\theta*}&=& \chi_*^2 \left[ a\ddot{a}+2(\dot{a}^2+k)\right]\,.\label{3rd-junc-cond-tangent-tangent}
\end{eqnarray}

The quantity $R^+_*\equiv R^+(t_*,r_*)$ and $R^+_{\mu\nu*}\equiv R^+_{\mu\nu}(t_*,r_*)$ are the Ricci scalar and the components of the Ricci  tensor on the outer side of star's surface. Therefore, the third junction contains three independent conditions.

\item \textbf{Fourth junction condition:} Using $[h_{ab}]=0$ and $[K_{ab}]=0,$ the fourth junction condition can be written as~\cite{Reina:2015gxa}
\begin{equation}
0=[\nabla_\rho R_{\mu\nu}]=n_\rho n^\sigma [\nabla_\sigma R_{\mu\nu}]+h^\sigma_\rho\nabla_\sigma [R_{\mu\nu}]\,.
\end{equation}

Using $[R_{\mu\nu}]=0$ the second contribution in the last equation vanishes and we can also show that~\cite{Reina:2015gxa}
\begin{eqnarray}
n^\sigma \left[ \nabla_\sigma R_{\mu\nu}\right]&=& h_\mu^\alpha h_\nu^\beta n^\sigma \left[\nabla_\sigma R_{\alpha\beta} \right] +n_\mu n_\nu n^\alpha n^\beta n^\sigma \left[ \nabla_\sigma R_{\alpha\beta}\right] \nonumber \\[1.5mm]
&=& h_\mu^\alpha h_\nu^\beta n^\sigma 
 \left[\nabla_\sigma R_{\alpha\beta} \right] + \frac{1}{2}n_\mu n_\nu n^\sigma \left[\nabla_\sigma R\right]\,,
\end{eqnarray}
where to go from the first to the second line we used $n^\alpha n^\beta n^\sigma [\nabla_\sigma R_{\alpha\beta}]=\frac{1}{2}n^\sigma [\nabla_\sigma R].$  Therefore, using $[h_{ab}]=0$, $[K_{ab}]=0$ and $[R_{\mu\nu}]=0,$ the fourth junction condition $[\nabla_\rho R_{\mu\nu}]=0$ becomes equivalent to the following conditions:
\begin{eqnarray}
h^\mu_\alpha h^\nu_\beta n^\sigma [\nabla_\sigma R_{\mu\nu}]=0  & \Leftrightarrow &  e^\mu_a e^\nu_b n^\sigma [\nabla_\sigma R_{\mu\nu}]=0\,, \label{tangent-tangent-4th-junc}
\end{eqnarray}
and
\begin{equation}
n^\rho [\nabla_\rho R]=0   \label{normal-normal-4th-junc}\,.
\end{equation}
Note that there is no contribution from any tangent derivative as well as any tangent-normal component of the normal derivative. The explicit form of the scalar condition~\eqref{normal-normal-4th-junc} is
\begin{eqnarray}
\frac{\dot{r}_*}{A_*}\partial_r R^+_*+\frac{\dot{t}_*}{B_*}\partial_t R^+_*=0\,,\label{4th-junc-cond-norm-norm}
\end{eqnarray}
while the tangent-tangent components in eq.~\eqref{tangent-tangent-4th-junc} give
\begin{eqnarray}
\!\!\!\!\!\!\!\!A_*\left(r_*\partial_r R^+_{\theta\theta*}-2R^+_{\theta\theta*}\right)\dot{t}_*+r_*B_*\partial_t R^+_{\theta\theta*}\dot{r}_*&=&0\,, \qquad\\[3mm]
\!\!\!\!\!\!\!\!-\frac{1}{\left(A_* B_*\right)^{3/2}}\left[ 
A_*^2\dot{t}_*\left(\dot{r}_*R_{rr*}^++\dot{t}_*R_{tr*}^+ \right) \left(\dot{r}_*\partial_r B_*+\dot{t}_*\partial_t B_*\right)
\right.&&\nonumber\\[1.5mm]
-A_*^2B_*\dot{t}_*\left( \dot{r}_*^2\partial_r R_{rr*}^{+} +2\dot{r}_*\dot{t}_*\partial_r R_{tr*}^{+} +\dot{t}_*^2\partial_r R_{tt*}^+ \right) &&\nonumber\\[1.5mm]
+A_*B_* \left(\dot{t}_*\partial_r A_*+\dot{r}_*\partial_t B_*\right) \left(\dot{r}_*^2 R_{rr*}^++2\dot{r}_*\dot{t}_* R_{tr*}^+ + \dot{t}_*^2R_{tt*}^+ \right)  &&\nonumber\\[1.5mm]
-A_*B_*^2\dot{r}_*\left( \dot{r}_*^2\partial_t R_{rr*}^++2\dot{r}_*\dot{t}_*\partial_t R_{tr*}^{+} +\dot{t}_*^2\partial_t R_{tt*}^+ \right) && \nonumber\\[1.5mm]
\left.+B_*^2\dot{r}_*\left(\dot{t}_*R_{tt*}^{+}+\dot{r}_*R_{tr*}^+ \right) \left(\dot{r}_*\partial_r A_*+\dot{t}_*\partial_t A_*\right)
\right]&=&0\,,\qquad
\label{4th-junc-cond-tang-tang}
\end{eqnarray}
Therefore, the fourth junction contains three additional independent conditions.

\end{itemize}

In summary, we have shown that in Quadratic Gravity, i.e. for the gravitational action~\eqref{action-weyl->Ricci} with $\alpha\neq 0$ and $\beta\neq 0,$ we obtain additional junction conditions. We have 2 conditions coming from the first, 2 from the second, 3 from the third and 3 from the fourth junction condition, which in general give a total of 10 independent matching conditions. 

To find the exact exterior metric we have to solve the Quadratic Gravity field equations~\eqref{eom} for the ansatz~\eqref{ext-ansatz-t-r} compatibly with the set of junction conditions. This task is not trivial and in this work we do not do it. However, we can still find some interesting constraints on the exterior metric by combining some of the properties of the field equations and the junction conditions.

\subsection{No-go theorems}

We now derive some no-go theorems that can be used to constrain the form of the metric functions $A(t,r)$ and $B(t,r)$. These were already derived in the context of $f(R)$ theories in Ref.~\cite{Senovilla:2013vra,Casado-Turrion:2022xkl}. Here we will adapt the theorems to the case of Quadratic Gravity.

First of all, since we are interested in finding constraints on $A(t,r)$ and $B(t,r)$ it is useful to make their dependence in the relevant equations more explicit. This can be done by eliminating $\dot{t}_*$ and $\dot{r}_*$ by means of one of the first and one of the second junction conditions, i.e. $A_*\dot{t}_*^2-B_*\dot{r}_*^2=1$ and $\beta=\beta_0\sqrt{A_* B_*}$, respectively. By doing so, we obtain
\begin{equation}
\dot{t}_*=\beta_0\sqrt{\frac{B_*}{A_*}}\,,\qquad \dot{r}_*=\sqrt{\beta_0^2-\frac{1}{B_*}}\,.
\label{rdot-tdot}
\end{equation}

We can now substitute~\eqref{rdot-tdot} into the second equation in~\eqref{2nd-junc-cond} and in the scalar component of the fourth junction condition in~\eqref{4th-junc-cond-norm-norm}:
\begin{eqnarray}
&&\dot{\beta}=\frac{1}{2}\left[ \beta_0^2\left(\frac{B_*}{A_*}\partial_t A_*-\partial_t B_*\right)+\frac{1}{B_*}\partial_t B_* \right]\,,\label{rdot-tdot-->junctions-2}
\\[2mm]
&& \sqrt{\beta_0^2-\frac{1}{B_*}}\partial_r R^+_*+\sqrt{\frac{A_*}{B_*}}\beta_0\partial_t R^+_*=0\,.
\label{rdot-tdot-->junctions}
\end{eqnarray}

These equations hold on the stellar surface only, i.e. at $t=t_*(\tau)$ and $r=r_*(\tau).$ However, in a dynamical scenario (e.g. in a collapse) star's surface will not be static, namely both $t_*$ and $r_*$ will vary. 
This means that we can consider the above equations and all other junction conditions to be satisfied for generic $t$ and $r$, in such a way that they correspond to a set of partial differential equations for the unknowns $A(t,r)$ and $B(t,r).$ 

We now prove the following no-go theorems.

\paragraph{No-go theorem 1.} \textit{No spherically symmetric exterior spacetime of the form~\eqref{ext-ansatz-t-r} whose Ricci scalar is constant can be smoothly matched to an FLRW interior with uniform dust density in Quadratic gravity.}

\begin{proof}

By assumption $R^+=\text{const.}$ (i.e. $\partial_t R^+=0=\partial_r R^+$). Using the third junction condition $[R]=0$ it follows that $R^-=R^+=\text{const.}$ Then, we also have $\dot{R}^-=0$.

Now consider the trace equation~\eqref{trace-eom} for the interior metric:
\begin{eqnarray}
-\frac{1}{8\pi G_{\rm N}}R^-+\alpha \Box R^-=-\rho(\tau)\qquad \Leftrightarrow \qquad  \frac{1}{8\pi G_{\rm N}}R^-+\alpha \ddot{R}^-+3\alpha \frac{\dot{a}}{a}\dot{R}^-=\rho(\tau)\,.
\end{eqnarray}
Since $\dot{R}^-=0$, the trace equation becomes
\begin{eqnarray}
 \frac{1}{8\pi G_{\rm N}}R^-=\rho(\tau)\,.
\end{eqnarray}
Then using $\rho(\tau)=\rho_0 /a^3(\tau),$ we can write
\begin{eqnarray}
a^3(\tau)=\frac{8\pi G_{\rm N}\rho_0}{R^-}=\text{const.}\,,
\end{eqnarray}
from which it follows  $a(\tau)=\text{const.},$ that implies $r_*(\tau)=a(\tau)\chi_*=\text{const.}$ However, this is inconsistent with a dynamical scenario as the radius of the star $r_*(\tau)$ cannot be constant. 

Therefore, we conclude that the assumption of constant Ricci scalar is inconsistent with an FLRW interior with uniform dust density in the case of the spherically symmetric exterior spacetime~\eqref{ext-ansatz-t-r}.

\end{proof}

\paragraph{No-go theorem 2.} \textit{No spherically symmetric exterior spacetime of the form~\eqref{ext-ansatz-t-r} can be smoothly matched to an FLRW interior with  uniform dust density in Quadratic Gravity if $R^+$ is either a function of $t$ only or of $r$ only.}

\begin{proof}

This statement follows from the previous no-go theorem and the fourth junction condition in eq.~\eqref{rdot-tdot-->junctions}.

\begin{itemize}

\item If $R^+=R^+(r),$ then the junction~\eqref{rdot-tdot-->junctions} reduces to $\sqrt{\beta_0^2-1/B}\partial_r R^+=0.$ This last equation is satisfied when at least one of the two factors vanishes. Note that $\beta_0^2=1/B$ would imply $\dot{r}=0$ which is inconsistent with a dynamical scenario. So the previous equation is satisfied for $\partial_r R^+=0$ which implies $R^+=\text{const.}$ But from the previous no-go theorem we know that this also leads to an inconsistency.

\item If $R^+=R^+(t),$ then eq.~\eqref{rdot-tdot-->junctions} reduces to $\beta_0\sqrt{A/B}\partial_t R^+=0.$ This last equation is satisfied when $\partial_t R^+=0$ which implies $R^+=\text{const.}$ But from the previous no-go theorem we know that this leads to an inconsistency.

\end{itemize}

\end{proof}

\paragraph{No-go theorem 3.} \textit{No spherically symmetric static exterior spacetime can be smoothly matched to an FLRW interior with uniform dust density in Quadratic Gravity.}

\begin{proof}

Consider $A=A(r)$ and $B=B(r)$ in the exterior metric~\eqref{ext-ansatz-t-r}, i.e. $\partial_t A=0=\partial_t B.$ This  implies that $\partial_t R^+=0.$ Then, from the junction~\eqref{rdot-tdot-->junctions} it follows that $\partial_r R^+=0.$ But from the no-go theorem 2 we know that this leads to an inconsistency.

Therefore, we conclude that the assumption $\partial_t A=0=\partial_t B$ is inconsistent with a dynamical FLRW interior configuration.

\end{proof}

\paragraph{No-go theorem 4.} \textit{No spherically symmetric exterior spacetime with single metric component  $B(t,r)=A^{-1}(t,r)$ can be smoothly matched to an FLRW interior with uniform dust density in Quadratic Gravity.}

\begin{proof}

From the first condition in~\eqref{2nd-junc-cond} we obtain
\begin{equation}
\beta=\beta_0\sqrt{A B}=\beta_0\quad \Rightarrow\quad \dot{\beta}=0\,,
\end{equation}
which, together with eq.~\eqref{rdot-tdot-->junctions-2}, gives
\begin{equation}
\left(\frac{2\beta_0^2}{A}-1\right)\partial_t A=0\,.
\end{equation}
If $A=2\beta_0^2=\text{const.},$ then from~\eqref{rdot-tdot} it follows $\dot{r}^2=\beta_0^2-A=-\beta_0^2<0$ which leads to an inconsistency. If $\partial_t A=0,$ then we have $A=A(r)$ and $B=A^{-1}(r),$ but from the no-go theorem 3 we know that a static exterior spacetime cannot be smoothly matched to a dynamical interior configuration.
Therefore, we conclude that a spherically symmetric exterior spacetime with single metric component -- either static or not -- is incompatible with an FLRW interior with uniform dust density.

\end{proof}

\subsection{Discussion}

The no-go theorems derived above put strong constraints on the functions $A(t,r)$ and $B(t,r)$ evaluated at the star's surface in the case of an FLRW interior with uniform dust density. The no-go theorems 1 and 2 tell us that the Ricci scalar $R_*^+$ cannot be constant and must depend on both $t_*(\tau)$ and $r_*(\tau)$. The no-go theorem 3 tells us that both $A_*$ and $B_*$ must be time-dependent. The no-go theorem 4 tells us that $B_*\neq 1/A_*.$ These results imply that the known black hole metrics~\cite{Lu:2015cqa} cannot be smoothly matched to the interior solution for generic values 
of $t$ and $r$. Indeed, as briefly reviewed in sec.~\ref{sec:summ-static-sol}, the known black hole solutions in Quadratic Gravity are static and have zero Ricci scalar. 

Now we have two main possibilities. (i)~There exists no exterior solution that can be possibly matched to a collapsing star with uniform dust density and whose interior spacetime is homogeneous and isotropic. For example, it may happen that the interior geometry does not respect the symmetries of the matter configuration and FLRW is no longer a valid interior solution; or it may happen that an initially uniform density configuration dynamically develops a radial profile in Quadratic Gravity\footnote{It is worth mentioning that for a static $(0,0)$ solution sourced by a spherical shell of matter, a radius-dependent Weyl curvature can emerge in the shell interior, as well as in the exterior~\cite{Holdom:2016nek}. In principle, an analogue breaking of homogeneity in the interior geometry could be found in a dynamical scenario too, for example in the interior of a collapsing uniform ball of dust.}. (ii)~Uniform dust collapse with an FLRW metric is a valid interior solution and the non-stationary exterior metric will tend to a static black hole metric (possibly to one of two solutions that are known in the literature) at late times and for distances sufficiently larger than the stellar radius. 
The second possibility means that the non-stationary deviations must become negligible asymptotically. 

Although we do not yet have a complete proof of its validity, here we assume that a collapsing uniform ball of dust with FLRW spacetime interior can be smoothly matched to some non-stationary exterior geometry. Since we have shown that in this case a horizon forms, we then expect that such exterior solution will asymptotically tend to a static black hole metric.

The fact that the exterior solution will be non-stationary is actually a quite general situation that also occurs in GR. For example, if we consider a rotating collapsing star in GR, the Kerr metric cannot be smoothly matched for generic values of the coordinates due to the non-stationarity of the exact exterior solution~\cite{Stark:1985da,Abrahams:1993wa,Lehner:2011aa,Nathanail:2017wly}. However, the Kerr metric becomes the exterior vacuum solution in the regime of late times and distances larger than the radius of the star, where the non-stationary deviations become negligible. Although we are not considering a rotating scenario, our situation is similar: the exact exterior solution is non-stationary and a static black hole solution may be reached only at late times and in the region $r\gg r_*,$ where the non-stationary deviations tend to zero.\footnote{In GR the spherically symmetric case is very special due to Birkhoff's theorem which guarantees that the Schwarzschild metric is the unique vacuum solution that can be exactly and smoothly matched to a collapsing star with uniform dust density.}

As briefly reviewed in sec.~\ref{sec:summ-static-sol}, we know that the possible family of solutions that is interesting for us is $(s,t)_{r_0}=(-1,1)_{r_0}$ which contains both Schwarzschild and Schwarzschild-Bach black holes. 
However, the mass parameter of the Schwarzschild-Bach black hole cannot be arbitrarily large as it is bounded from above by $\mathcal{O}\left(M^2_{\rm p}/m_2\right)$~\cite{Lu:2015cqa}.  Current observations from torsion-balance experiments~\cite{Lee:2020zjt} give $m_2\gtrsim 10^{-30}M_{\rm p}$.  This is sufficiently stringent to rule out solar-mass Schwarzschild-Back black holes. Therefore, we expect that the standard Schwarzschild metric will still be the correct asymptotic solution for an astrophysical collapse. 

As a future work we intend to numerically solve the entire system of field equations and junction conditions to find the exact non-stationary exterior solution from which we will also be able to learn about the late-time and large-distance asymptotics.

\section{Conclusions}\label{sec:conclus}

In this paper we studied for the first time the gravitational collapse in Quadratic Gravity. We focused on the simplest scenario of a collapsing spherically symmetric star with uniform dust density whose interior geometry respects the symmetries of the matter configuration, i.e. homogeneity and isotropy. We divided the analysis into interior and exterior solutions. Let us summarize the main results.
\begin{itemize}

\item The assumptions of homogeneity and isotropy for the interior spacetime select an FLRW-type metric. This implies that the Weyl square term does not affect the interior solution. 

\item The interior field equations were solved numerically and it was shown that for physical choices of the initial conditions a singularity still forms and the collapse is faster as compared to the case of Oppenheimer-Snyder in GR.

\item We that the interior solution contains an apparent inner horizon. We showed its formation numerically by calculating the expansion parameters of the ingoing and outgoing vectors of a null congruence and verifying that they become both negative at some point during the collapse. The presence of an inner horizon also implies the existence of an outer horizon. This result is very important because it rules out horizonless metrics (such as 2-2 holes) as exterior solutions that can be matched to a collapsing uniform-density dust star whose interior geometry is of FLRW-type and leaves black holes as the only possibility.

\item The junction conditions were derived in the case of Quadratic Gravity. These were first obtained  in Ref.~\cite{Reina:2015gxa} in a general form. Here we specialized them to the case of proper matching for which no discontinuity is present on the stellar surface. We found six extra conditions in addition to the other four coming from the well-known Darmois-Israel junction conditions~\cite{Darmois1927,Israel:1966rt}.

\item Some no-go theorems were used to partly constrain the spacetime metric outside the star. Given the general spherically symmetric ansatz in eq.~\eqref{ext-ansatz-t-r} we found that the metric components evaluated at the stellar surface must be time dependent, have non-constant Ricci scalar that depends on both $r$ and $t,$ and must have two independent metric functions, i.e. $B(t,r)\neq 1/A(t,r).$ 

\item Despite the non-stationarity of the exact exterior solution, we explained that a static black hole metric can be reached in the regime of late times and distances larger than star's radius ($r\gg r_*$) where the non-stationary deviations become negligible. In particular, we argued that the static black hole metric that the exterior solution will tend to is the standard Schwarzschild one.  However, future investigations are necessary to determine the exact non-stationary exterior solution that could be possibly matched to a collapsing star with uniform dust density and FLRW interior. The important point to emphasize is that in this work we have introduced all the ingredients that are needed to tackle the problem, i.e. to solve the full system of field equations and junction conditions for the metric ansatz~\eqref{ext-ansatz-t-r}.

\end{itemize}

Do our results imply that horizons always form and that singularities cannot be avoided in Quadratic Gravity? 

We have studied the case of a homogeneous and isotropic interior geometry. While this is an interesting first step, it is still possible that an arbitrary amount of inhomogeneity or anisotropy could emerge and play an important role during the collapse. In fact, the conformal flatness of the FLRW interior metric has made the Weyl tensor irrelevant for the interior solution. However, it would contribute to the interior dynamics in the presence of inhomogeneities or anisotropies. Moreover, the Weyl tensor would also be non-vanishing in a rotating scenario. At least in the linearized regime we know that the ghostly nature of the  Weyl-squared term induces a repulsive contribution in the gravitational force. Therefore, it is extremely interesting to understand whether the Weyl term can influence the qualitative features of the collapse, in particular the endpoint. 

Furthermore, our analysis is purely classical and does not take into account any quantum effects. 
Higher curvature terms in the action are required by renormalizability, but we have treated them only classically. Finally, no quantum aspects of the matter sector have been considered. It is quite reasonable to expect that quantum effects become important at some point during the collapse, when the matter has contracted sufficiently. In this regime, several kinds of generalizations of our study can be considered. First, we can treat the matter sector quantum mechanically and get additional terms on the right-hand side of the field equations. Second, quantum-gravitational corrections introduce additional higher-curvature~\cite{Daas:2023axu,Daas:2024pxs} and nonlocal operators~\cite{Knorr:2022kqp,Koshelev:2024wfk} into the action that in principle could change the physics at the endpoint of the collapse in nontrivial ways. Terms of this type may appear due to loop corrections or to completely non-perturbative quantum-gravitational effects~\cite{Bosma:2019aiu,Eichhorn:2022bgu,Platania:2023srt,Pawlowski:2023dda}.

All these questions are very fundamental and will be part of future works.


\subsection*{Acknowledgements}

The authors are grateful to Ramiro Cayuso and Bob Holdom for comments. L.~B. thanks Antonio Panassiti for enlightening discussions and the Institute of Theoretical Physics at Charles University for the warm hospitality during the final stage of this work. L.~B. acknowledges financial support from the European Union’s Horizon 2020 research and innovation programme under the Marie Sklodowska-Curie Actions (grant agreement ID: 101106345-NLQG). F.~D.~F and I.~K. acknowledge financial support by grants PRIMUS/23/SCI/005 and UNCE24/SCI/016 from Charles University. F.~D.~F is also grateful to GA\v{C}R 23-07457S grant of the Czech Science Foundation. I.~K. is grateful to Robert \v{S}varc for his valuable input on this project.
The work of F.~S.\ is supported by the Dutch Black Hole Consortium (DBHC) (grant no.\ NWA.1292.19.202) of the research programme NWA which is (partly) financed by the Dutch Research Council (NWO).



\bibliographystyle{utphys}
\bibliography{References}


\end{document}